\documentclass[5p]{elsarticle}

\usepackage{hyperref}
\usepackage[latin1]{inputenc}
\usepackage{color}           
\usepackage{soul}            
\usepackage{amssymb}
\usepackage{amsfonts}
\usepackage{amsmath}
\usepackage{wasysym}
\usepackage{graphicx}
\usepackage{epsfig}
\usepackage{amsthm}
\usepackage{bm}
\usepackage[normalem]{ulem}

\begin{document}


\title{Quantum correlations in periodically driven spin chains: Revivals and steady-state properties}
\author{Utkarsh Mishra$^{1,2}$, R. Prabhu$^{3}$, and Debraj Rakshit$^{4,5}$}
\address{$^1$Asia Pacific Center for Theoretical Physics, Pohang, Gyeongbuk 790-784, Korea
 \\
$^2$Institute of Fundamental and Frontier Sciences, University of Electronic Science and Technology of China, Chengdu 610051, China
 \\
 $^3$Department of Physics, Indian Institute of Technology Dharwad, Dharwad-580011, Karnataka, India
 \\
$^4$ICFO - Institut de Ciencies Fotoniques, The Barcelona Institute of Science and Technology,
Av. Carl Friedrich Gauss 3, 08860 Castelldefels (Barcelona), Spain
\\
$^5$Institute of Physics, Polish Academy of Sciences, Aleja Lotnikow 32/46, PL-02668 Warszawa, Poland}





\begin{abstract}
We study the dynamics of microscopic quantum correlations, viz., bipartite entanglement and quantum discord between nearest neighbor sites, in Ising spin chain with a periodically varying external magnetic field along the transverse direction.  Quantum correlations exhibit periodic revivals with the driving cycles in the finite-size chain. 
The time of first revival is proportional to the system size and is inversely proportional to the maximum group velocity of Floquet quasi-particles. On the other hand, the local quantum correlations 
in the infinite chain may get saturated to non-zero values after a sufficiently large number of driving cycles. 
Moreover, we investigate the  convergence of local density matrices, from which the quantum correlations under study originate, towards the final steady-state density matrices as a function of driving cycles. We find that the geometric distance, $d$, between the reduced density matrices of non-equilibrium state and steady-state obeys a power-law scaling of the form  $d \sim n^{-B}$, where $n$ is the number of driving cycles and $B$ is the scaling exponent. The steady-state quantum correlations are studied as a function of time period of the driving field and are marked by the presence of prominent peaks in frequency domain. The steady-state features can be further understood by probing band structures of Floquet Hamiltonian  and purity of the bipartite state between nearest neighbor sites. 
Finally, we compare the steady-state values of the local quantum correlations with that of the canonical Gibbs ensemble and infer about their canonical ergodic properties.  
Moreover, we identify generic features in the ergodic properties depending upon the quantum phases of the initial state and the pathway of repeated driving that may be within the same quantum phase or across two different equilibrium phases.
 
\end{abstract}

\maketitle

\section{Introduction}
Entanglement \cite{ HHHH}, in particular, and quantum correlations \cite{MODIetal},  in general, have continued to gain enormous amount of interests due to their numerous applications in quantum information theory \cite{nielson_book} and many-body physics
\cite{amitduttabook}. 
Quantum correlations are key resources for various quantum information processing tasks \cite{DC} and communication protocols \cite{one-way-qc} in many-body systems. 
They provide new insights about the cooperative phenomena in many-body systems such as quantum  phases, e.g. in the context of symmetry breaking phases \cite{nielsen_pra} and topological phases \cite{topological-qp}. Particularly, quantum criticality and related aspects have been investigated in great detail using various tools borrowed from quantum information theory \cite{concurrence,qd}. Moreover, recent experimental advances have successfully demonstrated effective manipulation of quantum correlations in several physical substrates \cite{lewenstein-rev,coldatom,optical-lattice,solid_xy,NMR}. 

Dynamics of closed many-body quantum systems has been a subject of intense research in recent years. Study of isolated systems out-of-equilibrium brings new possibilities for exploring physical phenomena, which are, understandably, not within  the reach of equilibrium statistical mechanics, and at the same time, provides a unique way of perceiving the emergence of equilibrium statistical properties \cite{ksgrmp,eisert_nat_phy}. Quenching one or several parameters of a many-body system happens to be a common strategy for exploring non-equilibrium dynamics  \cite{amicoetal2004,bose2003,quenched_dynamics, simulation_dyn,ergo_group,
kastner2014,myservival}. Ideas originating from quantum information have been incorporated in order to study a wide range of topics in the context of dynamics of closed quantum systems, such as revival and collapse phenomena of  entanglement \cite{collapse_revival}, Kibble-Zurek mechanism \cite{kibble}, thermalization, many-body localization \cite{MBL}, and decoherence of a qubit attached to a spin chain \cite{Jafari}. A widely studied fundamental topic is ergodicity of local quantities that relates the relaxation of the local properties of the many-body systems to their corresponding equilibrium values \cite{Rigol, Girardeau}. Time evolution of the microscopic quantum correlations under sudden quenching and their ergodic properties have been addressed in various spin systems \cite{ergo_group,nonintegrableergo, ergo_group1,sabre}.

Periodically disturbing a few parameters of an underlying Hamiltonian provides an interesting route for monitoring the out-of-equilibrium dynamics in multiples of driving-period \cite{sabre, book}. Such evolution, which is called \textit{stroboscopic} dynamics, has attracted vast interests in recent years as it offers a possibility of generating an effective Hamiltonian whose properties might be different from the initial Hamiltonian \cite{book,dynamic-localization,Rudner,hopping,arti-gauge}. Generation and relaxation of entanglement entropy towards a steady state have been studied in the periodically driven integrable spin models \cite{krishnendu2016,fazio2016,Apollaro2016}. An interesting observation of these works is that the entanglement entropy in the long-time steady state does not attain values corresponding to the infinite temperature state and hence contradicts the common perception of heating-up associated with the repeated disturbance \cite{heating-pd}. Recently, quantum critical scaling under periodic driving in spin systems has been studied \cite{pd-two-site-ent, Lazarides}. 

An important motivation for renewed interests in integrable models, such as quantum spin chains, is due to their experimental pertinence in present time. Experiments with ultracold atoms trapped in optical lattice have entered a very advanced  state. Near-perfect isolation from its environment and precise control over the trapping geometries and the inter-atomic interaction strengths have established cold-atom systems as an ideal platform for studying quantum phenomena in numerous many-body systems \cite{coldatom,optical-lattice}. Moreover, there are several other promising physical substrates, where many-body dynamics can be investigated in a controlled manner \cite{solid_xy,NMR,lab_xy_ion,xxz-exp}. Particularly, in recent times there have been considerable efforts for realizing periodically driven many-body systems in experiments \cite{ Eckardt}. 
Considering both experimental
and theoretical importance, we investigate the dynamics of quantum correlations in the periodically driven quantum
Ising spin chain, which represents a paradigmatic example of integrable system.

In this paper, we study the dynamics of microscopic quantum correlations, i.e., the quantum correlation between two nearest neighbor sites measured by concurrence and quantum discord, in a  periodically varying transverse field Ising model. More specifically, starting from a close to the zero-temperature initial state, the evolution is generated by repeated application of the unitary operator to reach at the desired time, $t=n\tau$. Here $n$ denotes the number of applied pulses and $\tau$ is the time-period between successive pulses.
We first consider the finite-size chain and observe that the quantum correlations show periodic revivals with respect to the driving cycles, $n$. A further examination of the time of first revival, $T_{r}$, reveals that it occurs at time $T_{r}\approx\frac{N}{\mbox{Max}[|v_{g}|]}$, where $N$ is the linear system size and $v_{g}$ is the quasi-particle group velocity. The evolution, here, is governed by the Floquet Hamiltonian, and the quasi-particle generated in the dynamics are interpreted as Floquet quasi-particles \cite{Apollaro2016}.  Here we try to extend the analysis of the revivals of quantum correlation beyond sudden quenching as reported in Ref.~\cite{sabre}.
 
Next, we consider the Ising model in the thermodynamic limit and study the relaxation of the bipartite quantum correlations
between two nearest neighbor sites as a function of $n$ for different choices of driving frequencies, $\omega={2\pi}/{\tau}$.  
We find that both the concurrence and the quantum discord tend to saturate to steady-state values after a sufficient number of driving cycles. We also calculate the distance, $d$, between the density matrix after $n$ driving cycles and the density matrix corresponding to the steady state obtained by taking the asymptotic limit $n\to \infty$. The distance, which goes to zero in the asymptotic limit, obeys power law scaling of the form $d\sim n^{-B}$ with respect to $n$. The fitted data shows that  $d$ goes to zero with scaling exponent $B=1.5$ for $\tau<2$ and $B=0.5$ for $\tau>2$.  

Next, for various choices of initial states, we investigate steady-state ($n\to \infty$) quantum correlations and study their variations as a function of $\tau$. The steady state quantum correlations are characterized by the presence of sharp peaks or kinks, which can be further understood by probing the band structures of the Floquet Hamiltonian, $H_{k,F}$. 

Since the local quantum correlations in the long-time steady-state may survive with finite values, it is interesting to check if the final steady-state value corresponds to a canonical Gibbs ensemble. If such canonical Gibbs states  exist, the quantity is termed as canonical ergodic. We consider two distinct cases depending on the choice of the driving pathway -- (i) repeated driving across the critical point, and (ii) repeated driving within a single phase. For case (i), when the initial state is chosen from the ordered phase, the quantum correlations always remain ergodic. On the contrary, and more interestingly, the quantum correlations may undergo ergodic to non-ergodic transitions in the frequency domain if the system is initialized in the disordered phase. For case (ii), we find situations, where the concurrence exhibits completely different ergodic behavior than the quantum discord. For such cases, although the quantum discord is characterized by ergodic to non-ergodic transitions in the frequency domain, the concurrence always remains ergodic when the driving is within the disordered phase.  We emphasize here that the above results on the canonical ergodicity under periodic driving show rich phenomena as compared to the previously observed results under single sudden quenching in the $XY$ model \cite{ergo_group}. For example, in the previous literature, quantum entanglement has been observed to be ergodic. However, non-ergodic to ergodic transition was noticed in quantum discord \cite{ergo_group}. We find that entanglement can also exhibit non-ergodic to ergodic transition in the frequency domain depending on the phase of the initial Hamiltonian and the driving frequency.  


The paper is organized as follows: Section~\ref{sec:model} introduces the spin model under study and the driving protocol employed in this work.  Section \ref{sec:revival} discusses about the revivals of the entanglement in the finite-size Ising model under periodic driving. Section \ref{sec:temporal} examines relaxation of the bipartite quantum correlations as a function of driving cycles $n$ in the thermodynamic limit. 
The results for the steady-state quantum correlations as a function of the time period of the driving protocol are analyzed in Sec.~\ref{sec:steady-state}. Section~\ref{sec: ergodicty-qc-pd} investigates ergodic properties of the quantities under study. 
Finally, Sec.~\ref{sec:conclusion} concludes and renders future perspectives. Appendices A, B, and C provide the definition of the time-dependent bipartite density matrix, the details of Floquet Hamiltonian, the derivations of time-evolved correlation functions, and the definitions of the quantum correlation measures considered in this work.

\section{The Model}
\label{sec:model}

In this paper, we consider one-dimensional Ising model in presence of a squared pulse transverse magnetic field. For academic interest, we sketch out the methodology for evaluating the time-evolved density matrices in Appendix A for generalized scenarios with quantum $XY$ model, whose Hamiltonian is given by
\begin{eqnarray}
\label{eq:HXY}
\textsl{H}(t) &=& \sum_{i=1}^{N}\frac{J}{4} \left[ (1+\gamma) \sigma^x_i \sigma^x_{i+1} +  (1-\gamma)  \sigma^y_i \sigma^y_{i+1}\right] \nonumber\\
& &- \frac{1}{2}\sum_i h(t)\sigma^z_i,
\end{eqnarray}
where \(J\) is pairwise coupling strength between nearest-neighbor spins,  $h(t)$ is a time-dependent external transverse magnetic field,  $\gamma$ is the anisotropy parameter, and $\sigma^x_i$, $\sigma^y_i$, $\sigma^z_i$ are the Pauli's spin matrices at the $i^{th}$ site. Ising model corresponds to the case when $\gamma=1$. The periodic boundary condition, i.e., $\sigma_{N+1}=\sigma_{1}$, is considered. In time-independent case, the above model undergoes  a quantum phase transition at $\lambda = 1$, where $\lambda = h/J$. For $\lambda>1$, the system is in paramagnetic or disordered phase and for $\lambda<1$, the system is in antiferromagnetic or ordered phase. The quantum criticality of the model has been widely studied via various approaches \cite{nielsen_pra,Lieb}. 
We consider a non-equilibrium scenario, where the system is driven periodically by introducing the time-dependent transverse magnetic field.   This magnetic field is taken in the form of a square-pulse, such as $h(t) = a$ for $t \le 0$ and for $t > 0$ 
\begin{equation}
 h(t) = \left\{ \begin{array}{ll}
         a & \mbox{if $(n-1)\tau \leq t \leq (n-\frac{1}{2})\tau$};\\
        b & \mbox{if $(n-\frac{1}{2})\tau \leq t \leq n\tau$}.\end{array} \right. \
        \label{eq:quenching}
 \end{equation}
Here $\tau (>0)$ is the time period between two successive pulses. 

The Hamiltonian in Eq.~(\ref{eq:HXY}) is exactly solvable even in the case of periodic driving. The details are presented in \ref{appc}. 
 The important operator in the Floquet formalism is the evolution operator, $U(\tau,a,b)$, over a period $\tau$. For the many-body quantum system under consideration, $U(\tau,a,b)=\otimes_{k}U_{k}(\tau,a,b)$ and the evolved state can be obtained as $\rho_{k}(n\tau)=\left[U_{k}(\tau,a,b)\right]^n\rho_{k}(0)\left[U{^{\dagger}}_{k}(\tau,a,b)\right]^n$. Here the initial state, denoted as $\rho_{k}(0 )$, is a thermal equilibrium state at $t=0$ and the unitary operator for the $k^{th}$ subspace is given by 
\begin{eqnarray}
U_k(\tau,a,b)&=&\exp[-iH_{k,F}\tau].
\label{eq:floquet-unt}
\end{eqnarray} 
Here $H_{k,F}$ is the Floquet Hamiltonian of the $k^{\text{th}}$ subspace.  
From the time dependent density matrix, $\rho_{k}(n\tau)$, one can calculate the reduced density matrix between two-sites, $\rho_{i,j}(n)$, as a function of driving cycles, $n$, in real space (defined in \ref{sec:densitymatrix}), which can further be expressed in terms of the two-point correlation functions derived in \ref{appc}. From the two-site reduced density matrix the local quantities under study, such as concurrence and quantum discord, are calculated. The variation of these quantities are examined with respect to time.  At this point, we would like to mention that when the time period $\tau$ of the driving protocol is small, i.e., the driving frequency is large, the time-evolution under periodic driving at the stroboscopic time and the one with single sudden quenching can be thought of representing equivalent dynamics \cite{ksgrmp}.  In this case, the post-quench Hamiltonian can be obtained by time-averaged Hamiltonian over one cycle of the periodic drive.  However, for an arbitrary driving period $\tau$, this equivalence does not hold.

\section{Finite system entanglement dynamics}
\label{sec:revival}
\begin{figure}
\includegraphics[width=0.5\textwidth]{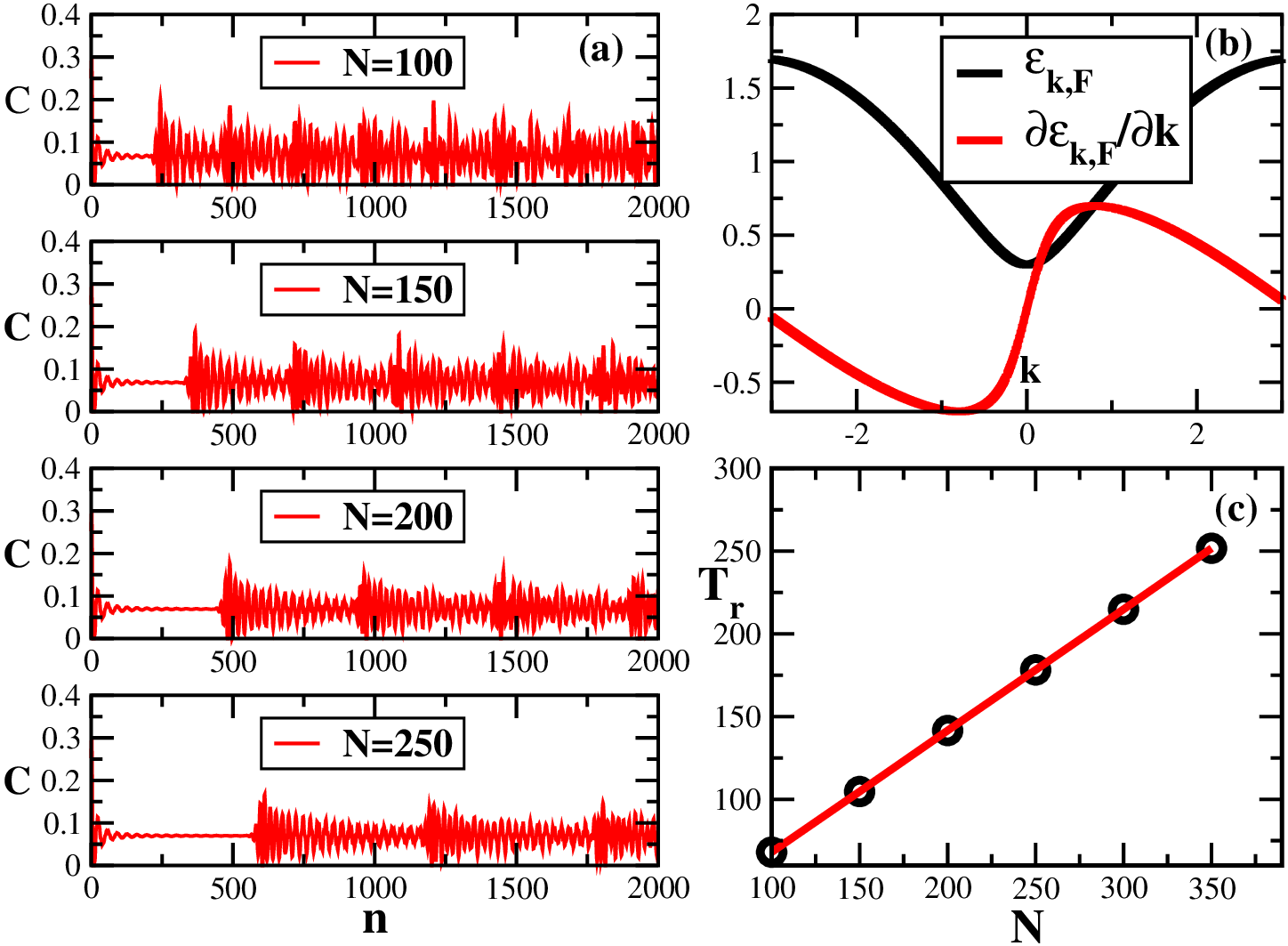}
\caption{(Color online.) (a) Evolution of concurrence with respect to driving cycles, $n$, for various system sizes, $N$. (b) The black and the red solid lines show the Floquet spectrum, $|\epsilon_{k,F}|$, and the group velocity, $v_{g}=|\frac{\partial \epsilon_{k,F}}{\partial k}|$, respectively. (c) The revival times, $T_{r}$, 
of concurrence are shown by black circles. The red line denotes the predicted revival time using $T_{r}\approx N/(2\mbox{max}(v_{g}))$. Here  $\tau=0.3$, $a/J=1.4$ and $b/J=0.0$.
 }
\label{fig:revival}
\end{figure}

In this section, we discuss the evolution of the quantum correlations in finite-size periodically driven Ising model. The initial states considered in this work are close to zero-temperature canonical equilibrium states of the system. Here we set $J \beta = 20$, where the temperature $T=1/k_{B} \beta$ and $k_{B}$ is the Boltzmann constant. In Fig.~\ref{fig:revival}, we show the dynamics of the concurrence as a function of $n$ for varying system size. The concurrence starts from its equilibrium value, and then oscillates around a mean value. The amplitude of the oscillation gradually reduces with an increase of $n$. After a finite number of driving cycles, $n_{rev}$, the concurrence again starts reviving followed by oscillations with comparatively larger amplitudes.
Such revivals have been reported in earlier studies with finite size $XY$ spin chain under sudden quench and time varying magnetic fields \cite{sabre,happola,Apollaro2016}. We find that the revival time, $T_{rev}=n_{rev}\tau$, changes with the system size (see Fig.~\ref{fig:revival}).  $T_{rev}$ turns out to be proportional to the system size and inversely proportional to the group velocity, $v_g=|\partial \epsilon_{k,F}/\partial k|$, where $\epsilon_{k,F}$ is Floquet spectrum (see  \ref{appc}). In Fig.~\ref{fig:revival}(b), the plot of the Floquet spectrum and the corresponding group velocity are shown. In Fig.~\ref{fig:revival}(c), we plot the revival time with respect to the system size. We find that the revival time follows the scaling $T_{r}\approx N/(2\mbox{max}(v_{g}))$.  This indicates that the revival time corresponds to the quasi-particles propagating with maximum group velocity. 

\section{Infinite system entanglement dynamics and steady states}
\label{sec:temporal}
After analysing the behavior of bipartite quantum correlations in finite-size Ising spin chain, we now consider them in the thermodynamic limit, $N\to \infty$. In the thermodynamic limit, we anticipate saturation of the long-time quantum correlations beyond certain driving cycles.  To confirm this, we consider the  periodic driving protocol on the external magnetic field (see Sec.~II). We study the temporal behavior of the bipartite quantum correlations as a function of  $n$ and monitor their convergence towards steady state values at long-time for different choices of $\tau$. 



\begin{figure}[t]
\center
\includegraphics[width=0.35\textwidth]{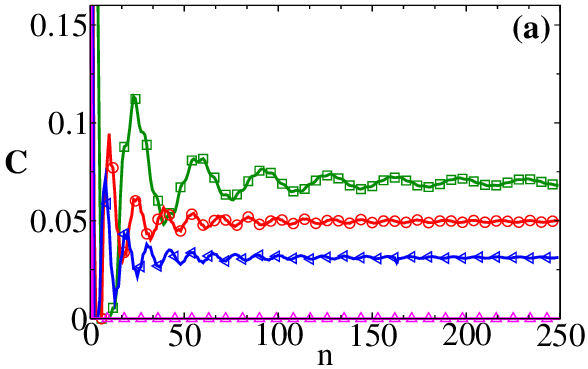}\vspace{0.2cm}
\includegraphics[width=0.35\textwidth]{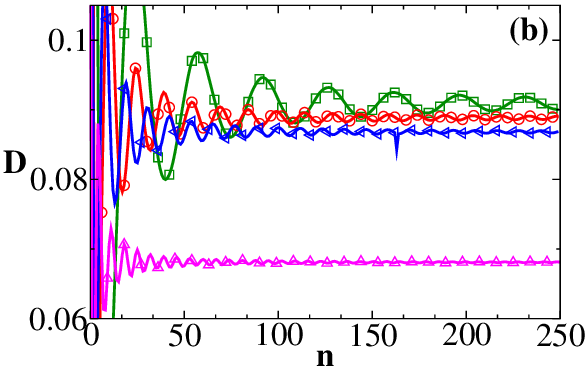}
\caption{(Color online.) (a) Concurrence, $C$, and (b) quantum discord, $D$, as function of driving cycle, $n$, of square pulsed magnetic field with $a/J=1.4$ and $b/J=0.0$ for different choices for $\tau$. The squares, circles, left-triangle, and up-triangles represent the cases with $J\tau/\hbar$=$0.3,0.7,0.9,$ and $1.5$, respectively. The solid lines serve as a guide to the eyes. Concurrence and discord are quantified in units of ebits and bits, respectively. 
 }
\label{fig:PD-ent-T}
\end{figure}


\begin{figure}[t]
\center
\includegraphics[width=0.23\textwidth]{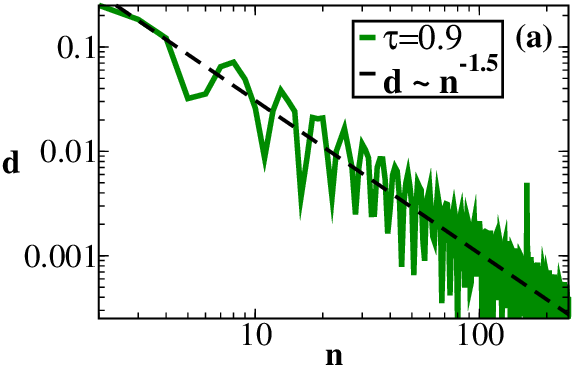}
\includegraphics[width=0.23\textwidth]{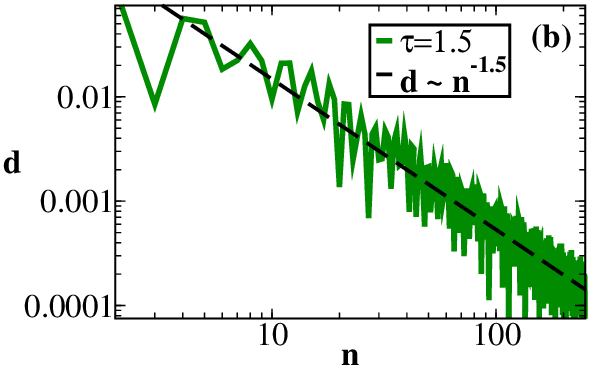}
\includegraphics[width=0.23\textwidth]{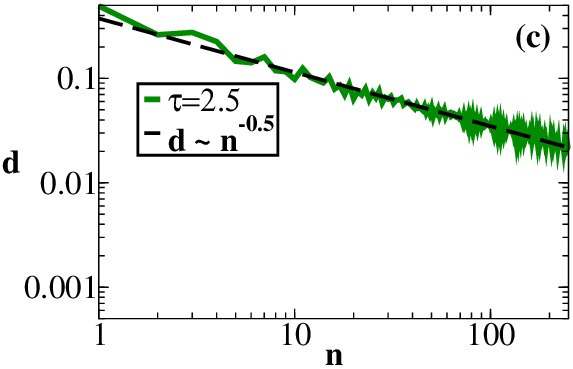}
\includegraphics[width=0.23\textwidth]{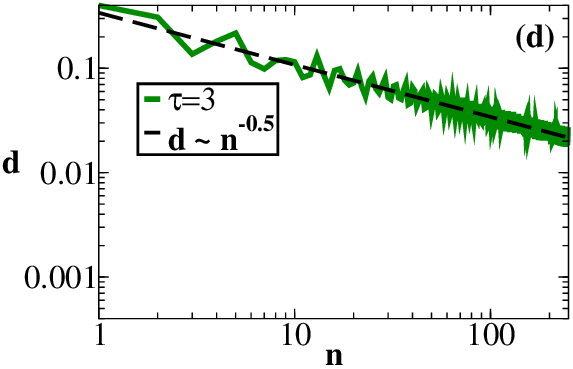}
\caption{(Color online.) Trace distance, $d$, as function of driving cycles, $n$, of square pulsed magnetic field with $a/J=1.4$ and $b/J=0.0$ for different choices of $\tau$.}
\label{fig:PD-dist-T}
\end{figure}

Here, we demonstrate the convergence of quantum correlations towards the steady-state values with driving cycles for system parameters, $a/J=1.4$ and $b/J=0$ for $0 < \tau \le 2.0$. The dynamics of the system is captured by the Floquet Hamiltonian, $H_{k,F}$, which is obtained by considering the dynamics at one complete driving period, $\tau$ (see Sec.~II and \ref{appc}). 

Figure \ref{fig:PD-ent-T} shows us the relaxations of (a) concurrence and (b) quantum discord for different values of $\tau$. In first few cycles, the quantum correlations oscillate with $n$ and then starts converging towards a steady-state value at large $n$. The saturated values depend upon the lengths of the period $\tau$. It can be seen from Fig.~\ref{fig:PD-ent-T} (a) that the entanglement saturates to lower values when $\tau$ is increased. The entanglement completely vanishes for $J\tau/\hbar=1.5$. Note that the survival of entanglement explicitly depends on the choice of $\tau$, and may again resurrect at  $\tau > 2.0$. A detailed analysis of the steady-state properties as a function of driving frequency has been carried out in the following section. 

Although a similar saturation behavior is observed for quantum discord (see Fig.~\ref{fig:PD-ent-T}(b)), unlike entanglement, quantum discord survives with finite value even at $J\tau/\hbar=1.5$. It is well known by now from several other works with sudden quenching \cite{ergo_group,myservival,collapse_revival,nonintegrableergo} that quantum discord is more robust against disturbance in comparison to entanglement, and long-time quantum discord usually survives even when entanglement vanishes. It is to be noted that the relaxation of bipartite quantum correlations in the finite-size $XY$ spin chain has been investigated under sinusoidal external magnetic field \cite{sabre}. 
Such form of the magnetic field, however, makes it hard to access the analytical form of the steady state.


An intuitive way of understanding the relaxation of various quantities, obtained from the time evolved density matrices, is to measure the distance between the density matrices itself.  In quantum information theory, such distances are often used and their properties have been studied in various contexts \cite{nielson_book}. For our case, we consider {\em trace distance}, $d$,  as a measure of overlap of information between two-density matrices. The  distance $d$ is defined as
\begin{equation}
d=\mbox{Tr}\sqrt{(\Delta \rho_n)^{\dagger} \Delta \rho_n} ,
\label{eq:distance}
\end{equation}
where $\Delta \rho_n=\rho_{12}(n)-\rho_{12}(\infty)$. Here $\rho_{12}(n)$ is the reduced density matrix of the bipartite state after $n$ driving cycles and $\rho_{12}(\infty)$ is the same in the limit of $n\to \infty$. Physically, Eq.~(\ref{eq:distance}) represents the distinguishability between two normalized density matrices. The maximum value, $d=1$, defines the maximum distinguishability between two states. In Fig.~\ref{fig:PD-dist-T}(a-d), the plot of distance  $d$ between the bipartite reduced density matrix at $t = \infty$ and the reduced density matrix of the driven system as a function of $n$ is shown. It is clearly seen from the plots that  $d$ approaches zero as $n$ increases. We also fit the data on power law function: $d = A n^{-B}$. The exponent $B$ suggests a qualitative change in the relaxation of the local quantities at $\tau=2$. We find the exponent $B$ is 1.5 for $\tau < 2$ and 0.5 for $\tau \ge 2$. It is worth mentioning here that the two possible dynamical phases depending on fast and slow periodic driving have been identified in \cite{krishnendu2016} while studying  the relaxation process of entanglement entropy.

\begin{figure}[t]
\includegraphics[width=0.23\textwidth]{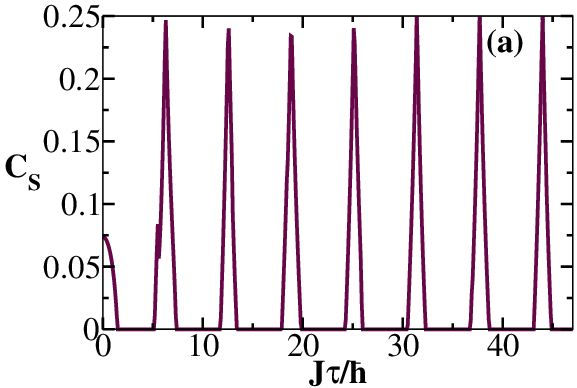}
\includegraphics[width=0.23\textwidth]{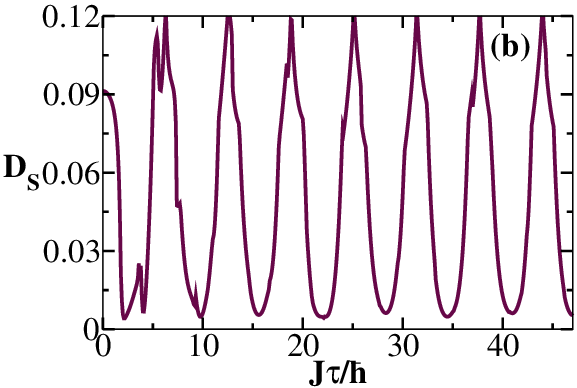}
\includegraphics[width=0.23\textwidth]{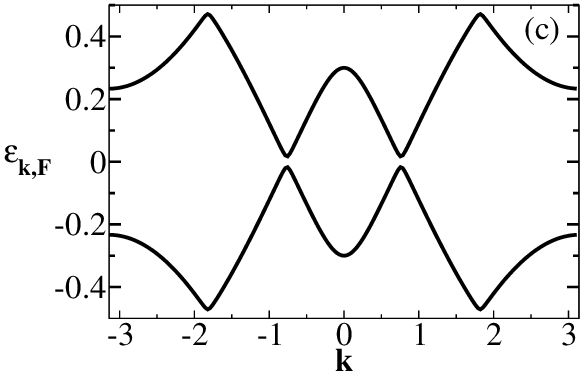}
\includegraphics[width=0.23\textwidth]{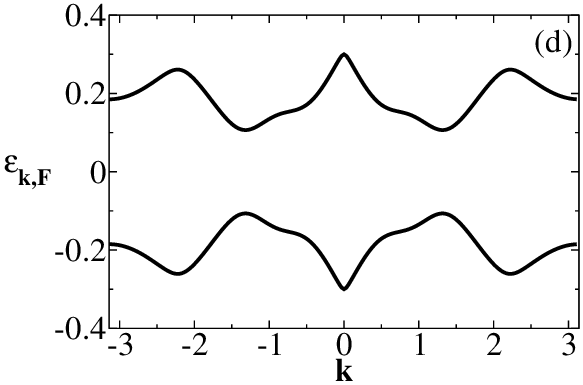}
\includegraphics[width=0.23\textwidth]{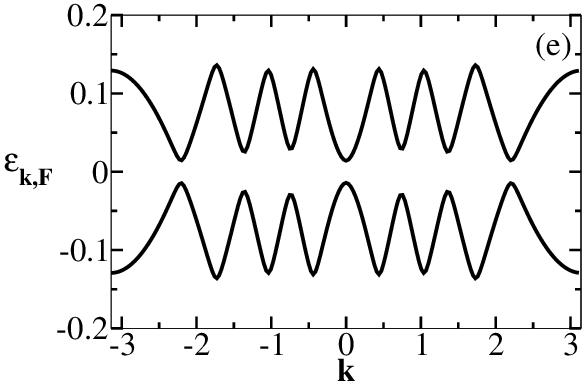}
\includegraphics[width=0.23\textwidth]{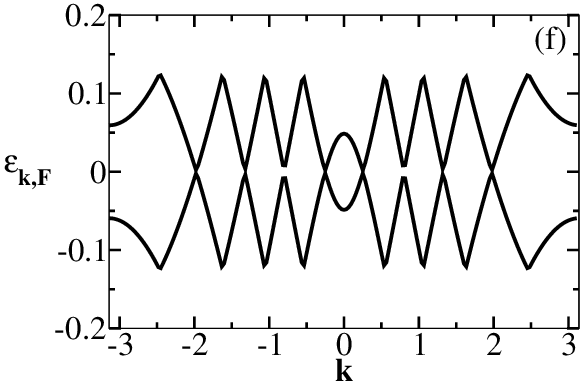}
\caption{(Color online.) Steady state quantum correlations and  Floquet spectrum for $a/J=1.4$ and $b/J=0.0$. (a) Steady-state concurrence, $C_{s}$, as a function of $\tau$. (b) Steady-state quantum discord, $D_{s}$, as a function of $\tau$. (c-f) Floquet spectrum, $\epsilon_{k,F}$, as a function of $k$ for $J\tau/\hbar=$ (c) 6.5, (d) 10, (e) 20 and (f) 25.} 
\label{fig:SS-PD-ent-T}
\end{figure}

\section{Quantum correlations in the long-time steady states}
\label{sec:steady-state}
In this section, we discuss the steady state behavior of the entanglement and the quantum discord as function of $\tau$. 
Figure \ref{fig:SS-PD-ent-T} shows steady-state entanglement (see Fig.~\ref{fig:SS-PD-ent-T}(a)) and quantum discord (see Fig.~\ref{fig:SS-PD-ent-T}(b)) as a function of the driving period for a particular case of squared pulse field with $a/J=1.4$ and $b/J=0$. Noticeably, repeated disturbance may not heat up the system indefinitely. Hence, the local quantum correlations present in the system may not be completely destroyed. There exist ranges of $\tau$, where the system possesses non-zero quantum correlations in the asymptotic limit. In fact, for the ranges of $\tau$, where the bipartite entanglement vanishes, the quantum discord survives with small but finite values.

Long-time steady-state quantum correlations show an oscillatory pattern in the frequency domain. They are characterized by the presence of sharp peaks. In low frequency domain, the peaks appear to be equispaced.  Our numerical data suggests that the peaks in the steady-state correlations may be a consequence of Floquet band crossing. Figure~\ref{fig:SS-PD-ent-T}(c-f) displays representative cases. Floquet spectrum, $\epsilon_{k,F}$, in momentum space, $k$, for $J\tau/\hbar=$ 6.5, 10, 20 and 25. It can be seen from Fig.~\ref{fig:SS-PD-ent-T}(a) and Fig.~\ref{fig:SS-PD-ent-T}(b) that  $J\tau/\hbar \approx 6.5$  and  $J\tau/\hbar \approx 25$ correspond to the kinks in quantum correlations, whereas $J\tau/\hbar=10$ corresponds to vanishing entanglement and minimum of quantum discord. For $J\tau/\hbar=20$, entanglement is zero but discord is finite valued (slightly higher than the minimum value). From Figs.~\ref{fig:SS-PD-ent-T}(c) and \ref{fig:SS-PD-ent-T}(f), it is perceptible that the peaks in quantum correlations are a consequence of Floquet band crossings. Quantum correlations assume minimum values for maximum energy gaps. We would like to mention here that appearance of the kinks in the frequency domain and their connection with  Floquet band crossings has been reported earlier in context of block entropy of the periodically driven systems \cite{krishnendu2016}.

We further analyzed these peaks in terms of the purity of the bipartite state $\rho_{AB}(n\to \infty)$.  It is worth mentioning that the state of the system at any time  $t=n\tau$ can be expressed in terms of  basis which are eigenvectors of $U_{k}(n\tau)$.  We note that two eigenvectors of $U_{k}(n\tau)$ become identical at the Floquet band crossing, implying reduced mixedness (or enhanced purity) of the state. A peak in the quantum correlation is a consequence of the same.

\begin{figure}[t]
\includegraphics[width=0.23\textwidth]{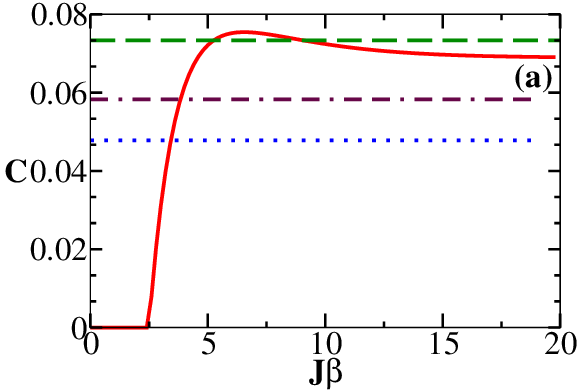}
\includegraphics[width=0.23\textwidth]{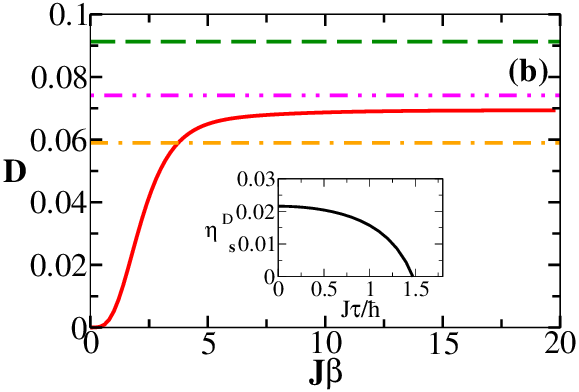}
\includegraphics[width=0.23\textwidth]{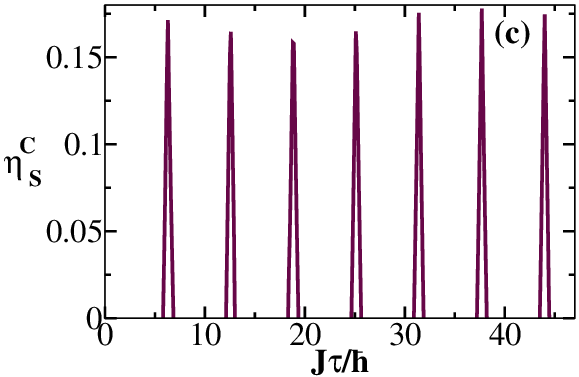}
\includegraphics[width=0.23\textwidth]{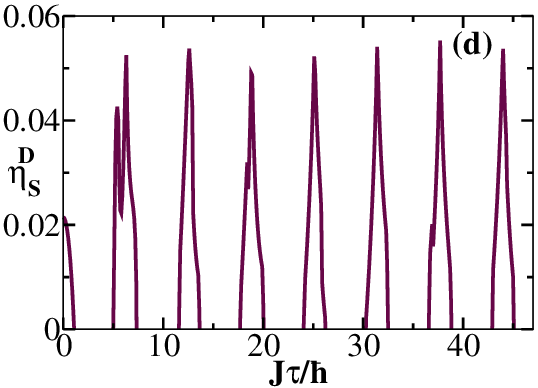}
\caption{(Color online.) {\it Periodic driving across phase transition -- disordered to ordered phase.} (a-b) Canonical equilibrium and steady-state quantum correlations for $a/J=1.4$ and $b/J=0.0$. (a) The solid line shows nearest-neighbor concurrence, $C$, of the canonical equilibrium state as a function of inverse temperature, $\beta$. The dashed, dash-dash-dotted, and dotted horizontal lines represent the steady-state concurrence for $J\tau/\hbar=$ 0.1, 0.8 and 1.0, respectively. (b) The solid line shows nearest-neighbor quantum discord, $D$, of the canonical equilibrium state as a function of inverse temperature, $\beta$. The dashed,  dash-dot-dotted, and dot-dash-dashed horizontal lines represent the steady-state quantum discord for $J\tau/\hbar=$ 0.1, 1.4, and 1.6, respectively. In the inset, the solid line shows ergodicity score of quantum discord, $\eta^D_s$, as a function of $\tau$.  (c-d) Ergodicity score of quantum correlations for the same set of parameters $a$ and $b$. (c) and (d), respectively, show ergodicity score of concurrence and quantum discord.} 
\label{fig:SS-ent-T}
\end{figure}

\section{Relaxation of quantum correlations under periodic driving}
\label{sec: ergodicty-qc-pd}
In the previous section, we discussed the behavior of the quantum correlations in the state at asymptotic limit $(n\to \infty)$. It would, therefore, be interesting to ask if the values of the evolved quantities correspond to the ones belonging to canonical thermal equilibrium states. In this section, we first describe the corresponding equilibrium state and the related notion of canonical ergodicity.

\subsection{Steady states and canonical ergodicity}
\label{subsec:ent_ev}
We consider that the spin chain is initially subjected to an external magnetic field $a$, and is in thermal equilibrium at temperature $T$ for $t \le 0$. The initial state, $\rho_{eq}(\beta,a)$, is given by $\exp[-\beta H(a)]$, where $\beta={1}/({k_{B}T})$ is the inverse of the absolute temperature $T$. The time evolved state, $\rho(\beta,a,b,t)$, at any time, $t>0$, is given by  $U^{\dagger}(a,b,t)\rho_{eq}(\beta,a)U(a,b,t)$. In the following, we discuss the ergodic properties within the notion of canonical equilibration, which we refer to as `canonical ergodicity'. Within this description, the ergodic properties of the system are inferred by comparing the time-evolved state at large time with the canonical equilibrium states \cite{ergo_group1,Lieb}. Note that ergodicity within generalized Gibbs ensemble \cite{LEV_email}, which is constructed by taking into account the conserved quantities, is not considered in this work.

\begin{figure}[t]
\includegraphics[width=0.4\textwidth]{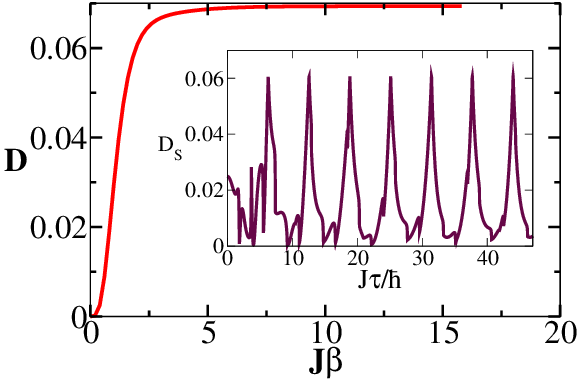}
\caption{(Color online.) {\it Periodic driving across phase transition -- ordered to disordered phase.} $a/J=0.0$ and $b/J=1.4$. The solid line shows canonical equilibrium quantum discord, $D$. In the inset, the solid line shows steady-state quantum discord, $D_S$, with respect to $\tau$ for the same set of parameters.}
\label{fig:ES-QD-PD-h1-0-h2-1o4}
\end{figure}

In order to construct a family of canonical equilibrium states suitable for describing the canonical ergodicity in periodic driving cases, we consider average Hamiltonian over one complete driving period \cite{sabre}, say $\bar{H} \equiv \frac{1}{\tau}\int_{0}^{\tau}H(t)dt$, and assemble the set of  canonical equilibrium states parametrized by $\tilde{\beta}$ for all $t>0$ as $\rho_{G}(\tilde{\beta},\bar{h}_{0})={\exp[-\tilde{\beta} H(\bar{h}_{0})]}$, where $\tilde{\beta}={1}/({k_{B}\tilde{T}})$ is the inverse temperature of the possible canonical equilibrium state and $\bar{h}_{0}$ is the average magnetic field over a period $\tau$ (See Eq.~(\ref{eq:quenching})). 
A quantity is termed ergodic, if the steady state value of the quantity, ${\cal Q}_{S}(\beta,a,b,\tau)$, approaches the ensemble average, ${\cal Q}_{G}(\tilde{\beta},\bar{h}_0)$, i.e.,
\begin{equation}
\label{erg-con}
\text{Tr} [\rho(\beta,a,b,\tau, n \to \infty) \hat{Q}] \approx \text{Tr} [\rho_{G}(\tilde{\beta},\bar{h}_{0})\hat{Q}].
\end{equation}
Alternatively, the value of a given observable in the steady-state limit simply mimics that of an equilibrium canonical ensemble at a given effective temperature $\tilde{\beta}$, which does depend on the observable.
 
If a local quantity does not satisfy Eq.~(\ref{erg-con}), it is called non-ergodic.
As we discuss below in the subsequent sections, the local quantum correlations may undergo ergodic to non-ergodic transitions. In order to monitor such transitions in a systematic manner, we use so called {\it ergodicity score} \cite{ergo_group}, which indicates the quantity's departure from ergodicity.  The ergodicity score, ${\cal \eta}_{S}^{{\cal Q}}$, is defined as
\begin{equation}
{\cal \eta}_{S}^{{\cal Q}}=\mbox {max}\Big[0,{\cal Q}_{S}(\beta,a,b,\tau)-\mbox{max}{\cal Q}_{G}(\tilde{\beta},\bar{h}_{0}) \Big].
\label{eq:ergodicityscore_pd}
\end{equation}
We search for all $\tilde{\beta}$ and choose the particular one for which ${\cal Q}_{G}(\tilde{\beta},\bar{h}_{0})$ attains maximum possible value for the given set of system parameters $a$, $b$.  This sets up an upper bound on the equilibrium value of ${\cal Q}_{G}(\tilde{\beta},\bar{h}_{0})$ as a function of $\tilde{\beta}$. This value is then compared with the corresponding steady state value of ${\cal Q}_{S}(\beta,a,b,\tau)$. Alternatively speaking, the difference between ${\cal Q}_{S}$ and ${\cal Q}_{G}$ is minimized.  A non-zero (vanishing) ergodicity score signals non-ergodic (ergodic) behavior of ${\cal Q}$. 

In the following sections, we look into possible situations that arise depending on the initial state and the choice of driving pathway.\\


\subsection{ Periodic driving across critical point.} 

Let us first look at the situation when the system is initiated in the disordered phase, and at each cycle, the driving Hamiltonian leads the system to a final state, which would correspond to the ordered phase of the system at equilibrium. For this case, let us use the parameters from the previous section, $a/J=1.4$ and $b/J=0$, which evidently belong to the case of repeated quenching of the system from disordered to ordered phase. The long-time steady-state values of bipartite entanglement and quantum discord for these chosen set of parameters have already been studied in Fig.~\ref{fig:PD-ent-T}. 

 Now, let us consider the cases for which  $J\tau/\hbar \le 2$. In Fig.~\ref{fig:SS-ent-T}(a), the red solid line shows the bipartite entanglement in the canonical equilibrium state,  as a function of inverse temperature $\beta$. The canonical equilibrium state is the one that correspond to the average magnetic field $\bar{h}_{0}$, i.e. $\rho_G(\beta,\bar{h}_0)$. It is clear from the figure that at high  temperature, i.e. at small $\beta$, the bipartite entanglement completely vanishes. However, below certain temperature, $J\beta \approx 2.2$, the system possesses finite amount of entanglement. At low-temperature ($J\beta\ge 10$), the entanglement saturates to a value close to $0.07$ {\it ebits} as the system approaches zero-temperature state. In order to find if the entanglement of the periodically driven system reaches to a state that corresponds to the canonical equilibrium state, we plot the entanglement of the long-time steady-state in the same figure. The steady-state values for varied $\tau$ are shown by horizontal lines in Fig.~\ref{fig:SS-ent-T}(a). We observe that the steady-state entanglement for all $\tau$ intersect the canonical entanglement at different temperatures, implying that entanglement is always ergodic in the frequency domain for this case. 

It is important to extend the analysis to {\it information-theoretic quantum correlation measures}, such as quantum discord, in order to encompass a more complete picture about the time-evolved density matrix. The definition of quantum discord is provided in  \ref{conc}. The red solid line in Fig.~\ref{fig:SS-ent-T}(b) shows behavior of the quantum discord in the canonical equilibrium state as a function of $\beta$. 
Quantum discord, unlike entanglement, shows monotonic behavior with respect to $\beta$.  It increases with decreasing system temperature and saturates at low enough temperature ($J\beta \geq 10$).  The steady state values of quantum discord for different $\tau$ are again shown by horizontal lines in the same plot. We find that for higher values of $\tau$, the steady-state quantum discord intersect the canonical equilibrium quantum discord curve at different temperature. However, surprisingly, below certain critical time-period, $\tau \le \tau_c$, there is no intersection, implying ergodic to non-ergodic transition of quantum discord in the frequency domain. For this case, we find $J\tau_c/\hbar \approx 1.5$. For clarity and an estimation of the degree to which the physical quantities under study is possibly non-ergodic, we calculate the ergodicity scores (see Eq.~(\ref{eq:ergodicityscore_pd})) both for entanglement, $\eta^{C}_{S}$, and quantum discord, $\eta^{D}_{S}$. In the inset of Fig.~\ref{fig:SS-ent-T}(b), we show  $\eta_S^D$ as function of $\tau$.  $\eta_S^D$ is finite valued for $J\tau/\hbar \le 1.5$, beyond which it becomes zero.  Ergodicity score for entanglement is zero throughout the range of $\tau$. 

However, the ergodic to non-ergodic transition in frequency domain is not unique to quantum discord. This becomes evident when the system is driven repeatedly with larger time-period. In fact, this information can be extracted from the steady-state values and canonical equilibrium values available in Figs.~\ref{fig:SS-PD-ent-T}(a-b) and \ref{fig:SS-ent-T}(a-b). The ergodicity scores for concurrence and quantum discord for this case are shown in Figs.~\ref{fig:SS-ent-T}(c) and \ref{fig:SS-ent-T}(d). The positive values of the ergodicity scores indicate that both entanglement and quantum discord become non-ergodic for that value of $\tau$.

 To find out if these features are generic, we perform similar analysis by initiating the system in the disordered phase but with different magnetic fields ($a/J>1$) and setting $b$ in the ordered phase $(0<b/J<1)$. We observe that for a given $a$, the quantum correlations undergo ergodic to non-ergodic transition  if $b<b_c$, beyond which both kinds of quantum correlations become ergodic irrespective of the driving frequency. For the case discussed above with $a/J=1.4$, we find $b_{c}/J\approx 0.8$. $b_c$ decreases with increasing $a$. 
When initial states are chosen from deep disordered phase, $b_c \to 0$, both entanglement and quantum discord become ergodic for any $b$ and $\tau$.\\
 
\begin{figure}[t]
\includegraphics[width=0.23\textwidth]{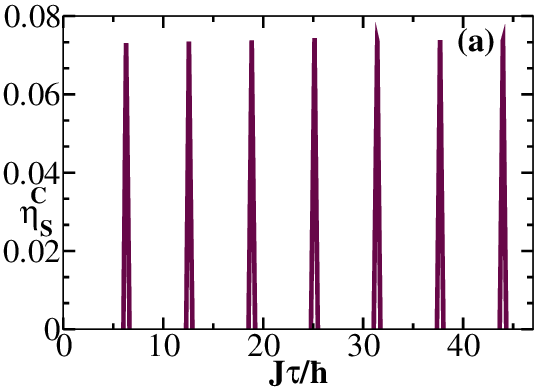}
\includegraphics[width=0.23\textwidth]{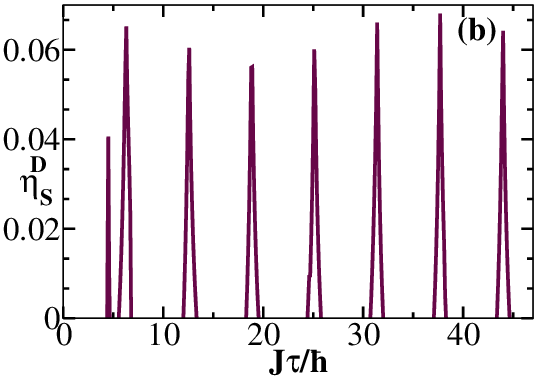}
\includegraphics[width=0.23\textwidth]{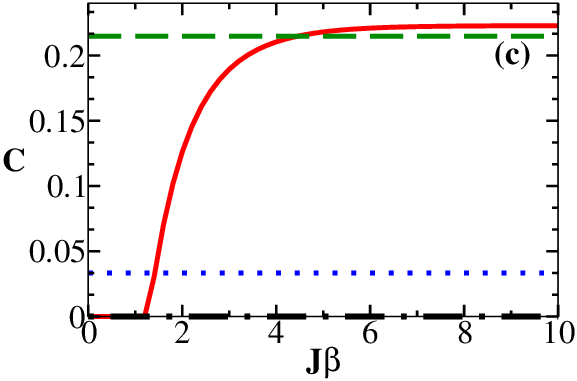}
\includegraphics[width=0.23\textwidth]{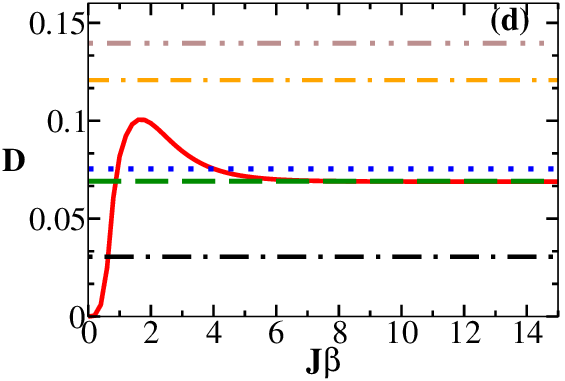}
\caption{(Color online.) {\it Periodic driving within same phase.} (a-b) Periodic driving within the ordered phase with $a/J=0.8$ and $b/J=0$. (a) and (b) shows quantum discord, $D$, and ergodicity score of quantum discord, $\eta_{S}^D$, respectively, as a function of $\tau$. (c-d) Periodic driving within the disordered phase with $a/J=2.4$ and $b/J=1.2$.  Canonical equilibrium and steady-state concurrence is shown in (c). The same for quantum discord is shown in (d). In (c-d), the solid lines represent the equilibrium values. In (c), the horizontal dashed, dash-dotted and dotted lines show steady-state values for $J\tau/\hbar$ = 0.1, 1.4 and 2.0, respectively. In (d), the horizontal dashed, dash-dotted, dotted, dash-dash-dotted and dash-dotted-dotted lines show steady-state values for $J\tau/\hbar$ = 0.1, 1.4, 2, 3, and 4, respectively.}
\label{fig:ES-QD-PD-h1o4-1o2}
\end{figure}
Let us now consider another possible scenario, where the system is driven from the ordered phase to the disordered phase at each driving cycle. The steady-state entanglement does not survive under this driving scheme. In fact, entanglement vanishes only after few cycles and stays so for any $n>0$. As a result, entanglement is trivially ergodic within this driving scheme. However, long-time steady-state quantum discord, as we may expect by now, survives, albeit with small value. Figure \ref{fig:ES-QD-PD-h1-0-h2-1o4} shows an example for quantum discord with $a/J=0$ and $b/J=1.4$ for canonical equilibrium state as a function of $\beta$. The inset of Fig.~\ref{fig:ES-QD-PD-h1-0-h2-1o4} shows steady-state quantum discord as a function of $\tau$.  It is easy to infer from the figure that steady-state values always correspond to certain thermal equilibrium state throughout the entire range of $\tau$. Hence, along with entanglement, quantum discord is also ergodic for this case.  Similar analysis is performed for arbitrary choices of $a$ and $b$ within this driving scheme. We find that both kinds of quantum correlations always remain ergodic.\\

\subsection{Periodic driving within same phase.} 
Finally, we discuss the system's response if repeated driving is conducted within the same phase. Let us first discuss the phenomenon when repeated driving is implemented within the ordered phase. For demonstration, a specific example is presented here with $a/J=0.8$ and $b/J=0$. Here, ergodic to non-ergodic transition is noticed in both kinds of quantum correlations (Figs.~\ref{fig:ES-QD-PD-h1o4-1o2}(a-b)). We investigate additional cases within this driving strategy, where time evolution starts from the same initial state ($a/J=0.8$) at $t=0$, and steady-state quantum correlations are studied by changing $b$. We find such transitions are not noticed when $0.8<b/J<1$ as both concurrence and quantum discord becomes ergodic in the entire frequency range.  Moreover, we find that irrespective of the initial states, quantum correlations behave ergodically whenever the system is driven close to the phase transition point.

Surprisingly, quantum correlation from entanglement-separability paradigm and  information-theoretic ones may not behave coherently when repeated driving is simulated within the disordered phase. In this case, although, concurrence always stays ergodic for arbitrary choices of $a$ and $b$, there may exist specific frequency windows, where quantum discord becomes non-ergodic, i.e., ergodic to non-ergodic transitions occur in frequency domain. Figure \ref{fig:ES-QD-PD-h1o4-1o2}  exhibits one such case with $a/J=2.4$ and $b/J=1.2$ for steady-state (c) entanglement and (d) quantum discord. However, although we always find a canonical equilibrium state that would correspond to the long-time steady-state concurrence for any given $\tau$ (cases for few specific choices of $\tau$ are shown by horizontal lines in Fig.~\ref{fig:ES-QD-PD-h1o4-1o2}(c)), the same is not true for quantum discord.

\section{Conclusion}
\label{sec:conclusion} 
In this work, we have investigated the behavior of microscopic quantum correlations, viz., bipartite entanglement and quantum discord, in a periodically driven Ising chain. Our finite-size analysis shows that the periodically driven quantum correlations exhibit periodic-revivals after a regular interval of driving cycles, $n$. The number of cycles (or time), after which such revival first occurs, can be predicted from the  Floquet quasi-particles picture and can be computed from the knowledge of Floquet spectrum alone. In infinite Ising chain, starting from an initial thermal equilibrium state, we observe that both entanglement and quantum discord eventually saturate after sufficient number of driving cycles. As may be expected, the required number of driving cycles for reaching saturation increases with increasing driving frequencies. 
Further understanding on dynamical relaxations are obtained by examining the trace distance between local density matrices of the driven system and the steady-state density matrix. The scaling of the distance measure $d$ shows a power law behavior, $d=A n^{-B}$, with respect to the driving cycles $n$. The exponent $B$, which turns out to be 1.5 or 0.5 depending upon the fast or slow periodic driving, indicates a qualitative change in the relaxation processes of the local quantities under study at $\tau=2$. Next, we study the steady-state quantum correlations with respect to $\tau$. Long-time steady-state quantum correlations  are characterized by the presence of prominent peaks in the frequency domain. Moreover, equipped by numerical evidences, we suggest a possible connection between the peaks and Floquet band crossings. 

Finally, we examine the canonical ergodicity of the quantum correlations under periodic driving. We find that the canonical ergodic properties of the quantum correlations crucially depend upon the quantum phase the initial state is chosen from, and the pathway of repeated driving. 
Particularly, within a repeated driving scheme via a square pulsed field, when an initial state is chosen from the disordered phase and the final field corresponds to the ordered phase, quantum correlations may display (canonical) ergodic to (canonical) non-ergodic transitions of the observables in the frequency domain.  The possible degree of non-ergodicity is indicated by so-called ergodicity score, $\eta^{Q}_{S}$, which shows that the observables oscillate between two possible situations, i.e., being ergodic or being non-ergodic, with the modulation of $\tau$. Moreover, for this case, we discuss conditions on the system parameters $a$ and $b$, for which such transitions appear.  For another choice of across the phase driving scheme, where the initial state belongs to the ordered phase and the final state belongs to the disordered phase, both the concurrence and the quantum discord turn out to be ergodic for any arbitrary driving frequency. Noticeably, when the initial states are chosen from ordered phase, i.e., $a \approx 0$, bipartite entanglement completely vanishes only after few driving cycles, although long-time quantum discord survives. 
Additionally, we discuss possibilities of ergodic to non-ergodic transitions in frequency domain, when the system is repeatedly driven within same phase. Surprisingly, the entanglement and the quantum discord behave differently in the entire frequency range if the driving is conducted within the disordered phase. In this case, we find that the entanglement remains ergodic for an arbitrary frequency of the square pulse. However, there exist frequency windows, where the quantum discord becomes non-ergodic.

Our work is relevant to current experimental set-ups for studying Floquet dynamics, particularly via ultracold atoms in optical lattice \cite{Eckardt}. Many interesting directions emerging from this work require independent attention. Particularly, the discussion on possible connection between Floquet band gap and peaks of the quantum correlations requires a rigorous study in order to achieve a better understanding. This is beyond the scope of the current work, and detailed attention will be paid in our future works. Other interesting questions exist. For example, how independent are the features on the choice of driving protocol? It will also be interesting to conduct analogous studies in non-integrable models and higher dimensional systems.

\section{Acknowledgments}
We acknowledge Lev Vidmar for useful comments. We thank Jhoanne Pedres Bon\~gol for careful reading and for helping us with the editorial aspects of the manuscript. RP acknowledges Science and Engineering Research Board, Government of India, for the financial support through Core Research Grant (Project File No. CRG/2018/004811). DR acknowledges the Spanish Ministry MINECO (National Plan 15 Grant: FISICATEAMO No. FIS2016-79508-P, SEVERO OCHOA No. SEV-2015-0522, FPI), European Social Fund, Fundacio' Cellex, Generalitat de Catalunya (AGAUR Grant No. 2017 SGR 1341 and CERCA/Program), ERC AdG OSYRIS, EU FETPRO QUIC, and the National Science Centre, Poland-Symfonia Grant No. 2016/ 20/W/ST4/00314. 
\appendix

\section{Density matrix of the system}
\label{sec:densitymatrix}
In order to study the microscopic quantum correlations, we need to find the corresponding density matrix of the system. 
%
%
%
The local density matrix of two-sites is given in terms of the single-site magnetization and two-site correlation functions as follows:

\begin{eqnarray}
\rho_{\ell m}(t) &=& \frac 14\big[\mathbb{I}_{\ell}\otimes \mathbb{I}_{m} \nonumber\\
&+& \sum_{\alpha=x,y,z} m^{\alpha}_{\ell}(\sigma^{\alpha}_{\ell} \otimes \mathbb{I}_{m})+ m^{\alpha}_{m}(\mathbb{I}_{\ell} \otimes \sigma^{\alpha}_{m}) \nonumber \\
&+ &  \sum_{\alpha,\, \beta=x,y,z}t_{\ell m}^{\alpha\beta}(\sigma_\ell^{\alpha} \otimes \sigma_m^{\alpha})\big],
\label{eq:dm12}
\end{eqnarray}
where $m^{\alpha}_{\ell}=\mbox{Tr}[\rho_{\ell} \sigma^{\alpha}_{\ell}]$ is the magnetization of the $i^{th}$ site along the $\alpha$-direction with corresponding single-site density matrix  $\rho_{\ell}=\frac 12 (\mathbb{I}+\vec m\cdot \vec\sigma)$, and $t_{\ell,m}^{\alpha\beta}=\mbox{Tr}[\rho_{\ell,m}(\sigma_{\ell}^{\alpha} \otimes \sigma_{m}^{\beta})]$ are the two-site spin-spin correlation functions given in \ref{appc}. Note that  we do not deal with symmetry broken ground state of the system in this work. Here the initial states are thermal states. Therefore, in the two-site density matrix, Eq.~\ref{eq:dm12}, $m^{x}$ and $m^{y}$ are identically zero as also discussed in Refs.~\cite{ergo_group, Lieb}. For a model with $Z_2$ symmetry broken ground state, on the other hand, it is known that the above two components of the magnetization will be non-zero \cite{refrepcit1,refrepcit2}. Moreover, translation symmetry of the system guarantees that $m^{\alpha}_{\ell}=m^{\alpha}_{m} $. 
Here $\mathbb{I}$ is the identity matrix in the Hilbert space of the single-site density matrix. In  \ref{appc}, we provide expressions for various correlators as a function of initial, final magnetic field and the temperature. These correlators can be used for constructing the two-site density matrices. 

\section{Floquet Hamiltonian and Correlation functions}
\label{appc}
In this Appendix, we outline the essential steps for obtaining the effective Floquet Hamiltonian and time-dependent spin-spin correlators in the $XY$ Hamiltonian (see Eq.~(1)).  We follow the route provided in reference~\cite{Lieb}. \\
 
\textit{Momentum space representation}: The first step is to write the lattice Hamiltonian, given in Eq.~(\ref{eq:HXY}), in $k$-space. 
We define the ladder operators, i.e., the raising, $a^{\dagger}_{i}$, and lowering, $a_{i}$, operators in terms of the spin operators as
\begin{eqnarray}
\sigma^{x}_{i}=a_{i}+a^{\dagger}_{i};~\sigma^{y}_{i}=-i(a^{\dagger}_{i}-a_{i});~  
\sigma^{z}_{i}=2a^{\dagger}_{i} a_{i}-1.
\label{eq:rl-op}
\end{eqnarray} 
 The operators $a_{j}$ and $a^{\dagger}_{j}$ are further written in terms of Fermi operators $b_{j}$ and $b^{\dagger}_{j}$   using the Jordan-Wigner transformation as 
 \begin{eqnarray}
 a_{j}=\exp(-i\pi\sum\limits_{\ell=1}^{j-1}b^{\dagger}_{\ell}b_{\ell})
b_{j};~a^{\dagger}_{j}=b^{\dagger}_{j}\exp(i\pi\sum\limits_{\ell=1}^{j-1}b^{\dagger}_{\ell}b_{\ell}).\nonumber\\ 
\label{eq:JW}
\end{eqnarray}
The next step is to Fourier transform the Fermi operators $b_{j}$ and $b^{\dagger}_{j}$ as:
\begin{eqnarray}
b_{j}(b^{\dagger}_{j})=\frac{1}{N}\sum\limits_{k=-N/2}^{k=N/2}\exp[\mp i j\phi_{k}]c_{k}(c^{\dagger}_{k}).
\label{eq:FT}
\end{eqnarray} 
where $\phi_{k}=\frac{2\pi k}{N}$. Using Eq.~(\ref{eq:rl-op}-\ref{eq:FT}) and related algebra on Eq.~(\ref{eq:HXY}), one  can obtain the Hamiltonian $H(t)=\sum_{k=1}^{N/2}\tilde{H}_{k}(t)$, 
where $\tilde{H}_{k}(t)$ is the Hamiltonian of the $k^{th}$ subspace given by
\begin{equation}
\tilde{H}_{k}(t)=\frac{1}{2}[\alpha(t)(c^{\dagger}_{k}c_{k}+c^{\dagger}_{-k}c_{-k})+i \delta_{k}(c^{\dagger}_{k}c^{\dagger}_{-k}+c_{k}c_{-k})+2 h(t)],
\label{eq:kspace-HXY}
\end{equation}
where $\alpha(t)=2[\cos\phi_{k}-h(t)]$, $\delta_{k}=-2\gamma \sin\phi_{k}$, and $c^{\dagger}_{k}$ and $c_{k}$ are the fermionic creation and annihilation operators in momentum space. 
In the chosen basis of the $k^{th}$ subspace \{ $|0,0\rangle, |k,-k\rangle, |k,0\rangle, |0,-k\rangle$\},  $\tilde{H}_{k}(t)$ can be expressed as a $4\times 4$ matrix:
\begin{equation}
\tilde{H}_{k}(t)=\left( \begin{array}{cccc}
h(t) & -\frac{i \delta_{k}}{2} & 0&0 \\
\frac{i \delta_{k}}{2} & 2 \cos \phi_{k}-h(t) & 0&0 \\
0& 0 & \cos \phi_{k} &0\\
0 & 0 & 0&\cos \phi_{k} \end{array} \right).
\label{eq:hm-dq}
\end{equation}\\

\textit{Floquet Hamiltonian:} 
The dynamics of the system under the periodic driving protocol described in Eq.~(\ref{eq:quenching}) is monitored via effective Floquet Hamiltonian. 
Noticing that the non-trivial contribution in system dynamics is originated from the reduced Hilbert space spanned by the basis $|0,0\rangle$ and $|k,-k\rangle$, it is sufficient to consider corresponding $2\times 2$ block of $\tilde{H_{k}}(t)$, which can be expanded in terms of the Pauli matrices as 
\begin{equation}
H_{k}(t)=c_{0}(k,t)\mathbb{I}_{2\times 2}+c_1(k)\sigma_{y}+c_{2}(k,t)\sigma_{z},
\label{eq:hm-pauli}
\end{equation}
where $c_{0}(k,t)$, $c_1(k)$, and $c_{2}(k,t)$ are  coefficients of the expansion defined as $c_{0}(k,t)=\cos \phi_{k}, c_{1}(k)=\gamma \sin\phi_{k},$ and $c_{2}(k,t)=-\cos \phi_{k}+h(t)$. 
We start the dynamics by assuming that the system is initially in a thermal equilibrium state (for all $t\le 0$). The corresponding equilibrium density matrix of the $k^{th}$ subspace is given by   
\begin{eqnarray}
\rho_{k}(0)=\exp[-\beta \tilde{H}_{k}(0)],
\label{eq:k-thstate}
\end{eqnarray}  
where  $\tilde{H}_{k}(0)$ is obtained by substituting $t=0$ in  Eq.~(\ref{eq:hm-dq}). 
Now we consider the periodic driving via external magnetic field $h(t)$ given in Eq.~(\ref{eq:quenching}). The  evolved state of the $k^{th}$ subspace after one complete driving period $\tau$ is given by
\begin{eqnarray}
\rho_{k}(\tau)=U_{k}(\tau)\rho_{k}(0)U_{k}^{\dagger}(\tau).
\end{eqnarray}
The unitary operator in one complete time period  $\tau$ is given by the time-order product of unitaries for each half-cycles:  
\begin{eqnarray}
U_k(\tau,a,b)&=&\exp \left[-i H_{k}(b)\frac{\tau}{2}\right]\exp \left[-i H_{k}(a)\frac{\tau}{2}\right]\nonumber\\
&=&\exp[-iH_{k,F}\tau],
\label{eq:floquet-unt}
\end{eqnarray} 
where  $H_{k}(a)= \vec{\sigma}.\vec{\epsilon}_{k}(a)=|\vec{\epsilon}_{k}(a)|\vec{\sigma}.\hat{n}_{k}(a); H_{k}(b)= \vec{\sigma}.\vec{\epsilon}_{k}(b)=|\vec{\epsilon}_{k}(b)|\vec{\sigma}.\hat{n}_{k}(b)$. Here $\hat{n}_{k}(a)=\frac{\vec{\epsilon}_{k}(a)}{|\vec{\epsilon}_{k}(a)|}$. The components of $\vec{\epsilon}_{k}(a)$ are given by $\vec{\epsilon}_{k}(a)=(0,c_{1}(k),c_{2}(k,a))$ where  $c_{2}(k,a)=-\frac{1}{2}(2\cos \phi_{k}-a)$. $\hat{n}_{k}(b)$ and its components are defined similarly.  The effective Floquet Hamiltoanin,  $H_{k,F}$, can also be written as $H_{k,F}=\vec{\sigma}.\vec{\epsilon}_{k,F}=|\vec{\epsilon}_{k,F}|\vec{\sigma}.\hat{n}_{k,F}$, where $\hat{n}_{k,F}=\frac{\vec{\epsilon}_{k,F}}{|\vec{\epsilon}_{k,F}|}$.  The quasi-energies are obtained as
\begin{eqnarray}
|\vec{\epsilon}_{k,F}|&=&\frac{1}{\tau}\mbox{Arccos}[\cos(|\vec{\epsilon}_{k}(b)|\frac{\tau}{2})\cos(|\vec{\epsilon}_{k}(a)|\frac{\tau}{2})]\nonumber\\
&-&\hat{n}_{k}(a).\hat{n}_{k}(b)\sin(|\vec{\epsilon}_{k}(b)|\frac{\tau}{2})\sin(|\vec{\epsilon}_{k}(a)|\frac{\tau}{2}),\nonumber\\
\label{eq:floquet-band}
\end{eqnarray}
and the $\hat{n}_{k,F}$ is given by
\begin{eqnarray}
\hat{n}_{k,F}&=&\frac{\hat{n}_{k}(b)}{{\cal N}_{k}}\sin(|\vec{\epsilon}_{k}(b)|\frac{\tau}{2})\cos(|\vec{\epsilon}_{k}(a)|\frac{\tau}{2})\nonumber\\
&+&\frac{\hat{n}_{k}(a)}{{\cal N}_{k}}\sin(|\vec{\epsilon}_{k}(a)|\frac{\tau}{2})\cos(|\vec{\epsilon}_{k}(b)|\frac{\tau}{2})\nonumber\\
&-&\frac{\hat{n}_{k}(b)\times \hat{n}_{k}(a)}{{\cal N}_{k}}\sin(|\vec{\epsilon}_{k}(b)|\frac{\tau}{2})\sin(|\vec{\epsilon}_{k}(a)|\frac{\tau}{2}),\nonumber\\
\end{eqnarray}
where ${\cal N}_{k}={|\vec{\epsilon}_{k,F}|\sqrt{1-\cos^{2}(|\vec{\epsilon}_{k,F}|)}}$.
Once the Floquet Hamiltonian is obtained, the state of the system after $n$ driving cycle is simply given by 
\begin{equation}
\rho_{k}(n\tau)= \exp[in H_{k,F} \tau]\rho_{k}(0)\exp[-i nH_{k,F} \tau].
\label{eq:time-dm-k}
\end{equation}\\

\textit{Spin-spin correlators:}
We now proceed to the derivation of the average magnetization, $m^{z}(n\tau)=\frac{1}{N}\sum\limits_{j=1}^{N}\langle\sigma^{z}_{j}
\rangle_{\rho(n\tau)}$, and spin-spin correlators,  $t^{\alpha\beta}_{i,j}(n\tau)=
\langle\sigma^{\alpha}_{i}
\sigma^{\beta}_{j}\rangle_{\rho(n\tau)}$, where the averages are performed on the time-dependent state $\rho(n\tau)$.  It is shown in \cite{Lieb} that the spin-spin correlation functions can be expressed in terms of fermionic operators $A'$s and $B'$s, where $A_{\ell}=c^{\dagger}_{\ell}+c_{\ell}$ and $B_{\ell}=c^{\dagger}_{\ell}-c_{\ell}$. Following this procedure for the case of periodic driving, it is straight forward to write $t^{\alpha\beta}_{i,j}(n\tau)$ as

 \begin{eqnarray}
t_{\ell, \ell+1}^{xx}(n\tau)&=&\langle B_{\ell}A_{\ell+1} \rangle_{\rho(n\tau)}\nonumber\\
t_{\ell, \ell+1}^{yy}(n\tau)&=&-\langle A_{\ell}B_{\ell+1} \rangle_{\rho(n\tau)}\nonumber\\
t_{\ell, \ell+1}^{zz}(n\tau)&=&\langle  A_{\ell}B_{\ell+1}\rangle_{\rho(n\tau)},\nonumber\\
t_{\ell, \ell+1}^{xy}(n\tau)&=&-i\langle  B_{\ell}B_{\ell+1}\rangle_{\rho(n\tau)}.
\label{eq:gcorr}
\end{eqnarray}\\
A further decomposition of the product of four fermionic operators of the form $\langle  A_{\ell}B_{\ell}A_{m}B_{m}\rangle$ via Wick's theorem eventually allows one to work with terms formed by the product of only two fermionic operators. In order to calculate the two-site correlators, we define
$G_{\ell,m}(n\tau)=\langle B_{\ell}A_{m}\rangle_{\rho(n\tau)},$
$G_{\ell,m}'(n\tau)=\langle A_{\ell}B_{m}\rangle_{\rho(n\tau)},
Q_{\ell,m}(n\tau)=\langle A_{\ell}A_{m}\rangle_{\rho(n\tau)}$, and 
$S_{\ell,m}(n\tau)=\langle B_{\ell}B_{m}\rangle_{\rho(n\tau)}$. Utilizing the definition of $A_{\ell}$ and $B_{\ell}$, we obtain
\begin{eqnarray}
G_{\ell,\ell+1}(n\tau)&=&\frac{1}{N}\sum\limits_{k=1}^{N/2}(-2 i\sin(\phi_{k}))\mbox{Tr}[X^{k}_{K'}\rho_{k}(n\tau)] \nonumber\\
&+&\frac{1}{N}\sum\limits_{k=1}^{N/2}(2\cos(\phi_{k}))\mbox{Tr}[m^{k}_{z}\rho_{k}(n\tau)],\nonumber\\
\label{eq:G}
\end{eqnarray}
\begin{eqnarray}
G'_{\ell,\ell+1}(n\tau)&=&\frac{1}{N}\sum\limits_{k=1}^{N/2}(-2 i\sin(\phi_{k}))\mbox{Tr}[X^{k}_{K'}\rho_{k}(n\tau)] \nonumber\\
&-&\frac{1}{N}\sum\limits_{k=1}^{N/2}(2\cos(\phi_{k}))\mbox{Tr}[m^{k}_{z}\rho_{k}(n\tau)],\nonumber\\
\label{eq:GP}
\end{eqnarray}
\begin{eqnarray}
Q_{\ell,\ell+1}(n\tau)&=&\frac{1}{N}\sum\limits_{k=1}^{N/2}(-2 i\sin(\phi_{k}))\mbox{Tr}[X^{k}_{K}\rho_{k}(n\tau)]\nonumber\\
&+&\frac{1}{N}\sum\limits_{k=1}^{N/2}2\cos(\phi_{k}),
\label{eq:Q}
\end{eqnarray}
\begin{eqnarray}
S_{\ell,\ell+1}(n\tau)&=&\frac{1}{N}\sum\limits_{k=1}^{N/2}(-2 i\sin(\phi_{k}))\mbox{Tr}[X^{k}_{K}\rho_{k}(n\tau)]\nonumber\\
&-&\frac{1}{N}\sum\limits_{k=1}^{N/2}2\cos(\phi_{k}),
\label{eq:S}
\end{eqnarray}
where $X^{k}_{K}=c^{\dagger}_{k}c^{\dagger}_{-k}-c_{k}c_{-k}$, $X^{k}_{K'}=c^{\dagger}_{k}c^{\dagger}_{-k}+c_{k}c_{-k}$, and $m^{k}_{z}=c^{\dagger}_{k}c_{k}+c^{\dagger}_{-k}c_{-k}-1$. 
Note that in the thermodynamic limit, $N\to \infty$, the summations in Eqs.~(\ref{eq:G}-\ref{eq:S}) can be replaced via integrals: $\frac{1}{N}\sum\limits_{k=1}^{N/2} \to \frac{1}{2\pi}\int_{0}^{\pi}d\phi $. From the expressions given in Eqs.~(\ref{eq:G}-\ref{eq:S}),  single-site and two-site quantities, required for constructing the two-body density matrices (see Eq.~(\ref{eq:dm12})), can be calculated. For example, the average magnetization  $m^{z}(\tau)$ can be obtain from $G_{\ell,m}(n\tau)$ as $m^{z}(n\tau)=\frac{1}{2}G_{\ell,\ell}(n\tau)$. For the nearest neighbor sites ($\ell,\ell+1$), the correlators are obtained as $t^{xx}_{\ell,\ell+1}(n\tau)=G_{\ell,\ell+1}(n\tau); ~t^{yy}_{\ell,\ell+1}(n\tau)=-G'_{\ell,\ell+1}(n\tau);~t^{zz}_{\ell,\ell+1}(n\tau)=m^{2}_{z}(n\tau)+G'_{\ell,\ell+1}(n\tau)G_{\ell,\ell+1}(n\tau)-Q_{\ell,\ell+1}(n\tau)S_{\ell,\ell+1}(n\tau);~t^{xy}_{\ell,\ell+1}(n\tau)=-Q_{\ell,\ell+1}(n\tau)$. Once the correlators after $n$ driving cycles are obtained, one can take the limit $n\to \infty$ in order to obtain steady-state two-site density matrix $\rho_{12}(\infty)$.

\section{Quantum correlation measures}
Here we provide the definitions for the quantum correlation measures -- concurrence \cite{concurrence}, as entanglement separability measure, and quantum discord, as information theoretic kind quantum correlation measure.


\label{conc}
\textit{Concurrence:} Concurrence  is a well known computable measure of entanglement of a bipartite quantum state in  $\mathbb{C}^{2}\otimes \mathbb{C}^{2}$. If $\rho_{AB}$ denotes  an arbitrary two-qubit quantum state, then its concurrence is given by 
C$(\rho_{AB})=\max\{0,\lambda_1-\lambda_2-\lambda_3-\lambda_4\}$,
where $\lambda$'s are square roots of the eigenvalues of $\widetilde {\rho}_{AB}\rho_{AB}$ in descending order with $\widetilde \rho_{AB} = (\sigma_{y}\otimes \sigma_{y})\rho_{AB}^{*} (\sigma_{y}\otimes \sigma_{y})$. Here $\sigma^{y}$ is the Pauli matrix and $\rho_{AB}^{*}$ is the complex conjugate of $\rho_{AB}$ in the same basis. Concurrence vanishes for separable states and attains unity for maximally entangled states.  In the figures labeling, we have denoted concurrence by $C\equiv C(\rho_{AB})$.  \\

\textit{Quantum Discord}: Quantum discord is a measure of quantum correlations beyond entanglement \cite{MODIetal,qd,ujjwal-review}. 
It utilizes the fact that the two equivalent definitions of the mutual information in terms of the classical probabilities are not same when their natural extensions are considered within quantum theory. 
The mutual information of the quantum state $\rho_{AB}$  given by ${\cal I} =S(\rho_{A})+S(\rho_{B})-S(\rho_{AB})$ defines total correlation of the state. Here $S(\rho_{A})$, $S(\rho_{B})$, and $S(\rho_{AB})$ are the von Neumann entropies defined as $S(\varrho)=-\mbox{Tr}(\varrho \log_2 \varrho)$. The quantity ${\cal I}$ can be interpreted as the amount of information shared by the two parties in a quantum state $\rho_{AB}$. 

The second quantum version of mutual information is given by ${\cal J}=S(\rho_{A})-S(\rho_{A|B})$, where $S(\rho_{A|B})$ is the conditional entropy, $S(\rho_{A|B}) = \min_{\{B_i\}}\sum_{i}p_{i}S(\rho_{A|i})$ and the measurement is performed on subsystem $B$ (in a similar way it can be defined for measurement on subsystem $A$). The measurement operators, $\{B_{i}\}$, are rank-1 projective operators and $p_{i}$'s are the probabilities obtained after the measurements on subsystem $B$. The measured state and the probability of output state are given by $\rho_{A|i}=\frac1p_{i} \mbox{Tr}_{B}[( I_{A}\otimes B_{i})\rho_{AB}(I_{A}\otimes B_{i})]$ and $p_i=\mbox{Tr}_{AB}[(I_{A}\otimes B_{i})\rho_{AB}(I_{A}\otimes B_{i})]$, respectively. We note that, in general, the minimization should be carried over all possible positive-operator valued measures (POVMs) in the definition of the quantum discord. However, it has been established that it is sufficient to consider only rank-1 projective measurements in most cases.

Once we have ${\cal I}$ and ${\cal J}$, the quantum discord is defined as $D={\cal I}-{\cal J}$, i.e., $D=S(\rho_{B})-S(\rho_{AB})+S(\rho_{A|B})$. Note that quantum discord reduces to the von Neumann entropy of the reduced state for pure bipartite states.\\


\section*{Bibliography}
\begin{thebibliography}{1}
\bibitem{HHHH} R. Horodecki, P. Horodecki, M. Horodecki, and K. Horodecki, Rev. Mod. Phys. \textbf{81}, 865  (2009).


\bibitem{MODIetal} K. Modi, A. Brodutch, H. Cable, T. Paterek, and V. Vedral, Rev. Mod. Phys. \textbf{84}, 1655 (2012).

\bibitem{nielson_book} M.A. Nielsen and I.L. Chuang, \textit{Quantum Computation and
Quantum Information} (Cambridge University Press, Cambridge, UK, 2000).

\bibitem{amitduttabook}  A. Dutta, G. Aeppli, B.K. Chakrabarti, U. Divakaran, T. F. Rosenbaum and D. Sen, \textit{Quantum Phase Transitions in Transverse Field Spin Models: From Statistical Physics to Quantum Information} (Cambridge University Press, Cambridge, 2015).

\bibitem{DC} C.H. Bennett and S.J. Wiesner, Phys. Rev. Lett. {\bf 69}, 2881 (1992);  C.H. Bennett, G. Brassard, C. Cr\' epeau, R. Jozsa, A. Peres, and W.K. Wootters, Phys. Rev. Lett. {\bf 70}, 1895 (1993); A. Ekert, Phys. Rev. Lett. \textbf{67}, 661 (1991); N. Gisin, G. Ribordy, W. Tittel, and H. Zbinden, Rev. Mod. Phys. {\bf 74}, 145 (2002).



\bibitem{one-way-qc} A. Sen(De) and U. Sen, Physics News, \textbf{40}, 17 (2010) (arXiv:1105.2412 [quant-ph]).

\bibitem{nielsen_pra} T.J. Osborne and M.A. Nielsen, Quantum Inf. Proc. {\bf 1}, 45 (2002); T.J. Osborne and M.A. Nielsen, Phys. Rev. A {\bf 66}, 032110 (2002); A. Osterloh, L. Amico, G. Falci, and R. Fazio, Nature {\bf 416}, 608 (2002);   L. Amico, R. Fazio, A. Osterloh, and V. Vedral, Rev. Mod. Phys. \textbf{80}, 517 (2008); U. Marzolino, S.M. Giampaolo, F. Illuminati, Phys. Rev. A \textbf{88}, 020301(R) (2013); T. Werlang, C. Trippe, G.A.P. Ribeiro, and G. Rigolin, Phys. Rev. Lett. \textbf{105}, 095702 (2010); M.S. Sarandy, Phys. Rev. A \textbf{80}, 022108 (2009); G. De Tomasi, S. Bera, J.H. Bardarson, and F. Pollmann, Phys. Rev. Lett. \textbf{118}, 016804 (2017); G. Vidal, Phys. Rev. Lett. \textbf{91}, 147902 (2003); \textbf{93}, 040502 (2004); J. Eisert, M. Cramer, M.B. Plenio, Rev. Mod. Phys. \textbf{82}, 277 (2010); F. Franchini, A.R. Its,  V. E. Korepin, and L. A. Takhtajan, Quant. Info. Proc. \textbf{10} 325 (2011); P. Zanardi, M. Cozzini, and P. Giorda, J. Stat. Mech. \textbf{2007}, 1742 (2007); A. Osterloh, G. Palacios, and S. Montangero, Phys. Rev. Lett. {\bf 97}, 257201 (2006); B. Tomasello, D. Rossini, A. Hamma and L. Amico, Europhys. Lett. {\bf 96}, 2 (2011).






\bibitem{topological-qp} C.J. Shan, W.-W. Cheng, J.-B. Liu, Y.S. Cheng, and T.-K. Liu, Sci. Rep. \textbf{4}, 4473 (2014);  Y.Xin Chen and S.W. Li, Phys. Rev. A \textbf{81}, 032120 (2010); S. Singha Roy, H. S. Dhar, D. Rakshit, A. Sen De, U. Sen,  J. Magn. Magn. Mater. 444, 227 (2017);  J. Cho and K.W. Kim, Sci. Rep. \textbf{7}, 2745 (2017); T.P. Oliveira and P. D. Sacramento, Phys. Rev. B \textbf{89}, 094512 (2014).









 

\bibitem{concurrence} S. Hill and W.K. Wootters, Phys. Rev. Lett. \textbf{78}, 5022 (1997); W.K. Wootters, {\em ibid.} \textbf{80}, 2245 (1998).

\bibitem{qd} H. Ollivier and W. H. Zurek, Phys. Rev. Lett. \textbf{88}, 017901 (2001); L. Henderson and V. Vedral, J. Phys. A: Math. Gen. \textbf{34}, 6899 (2001).

\bibitem{lewenstein-rev} M. Lewenstwein, A. Sanpera, V. Ahufinger, B. Damski, A. Sen(De), and U. Sen, Adv. Phys. \textbf{56}, 243 (2006); 

                          
\bibitem{coldatom} J. Struck, C. \"{O}lschl\"{a}ger, R. Le Targat, P.S. Panahi, A. Eckardt, M. Lewenstein, P. Windpassinger, and K. Sengstock, Science \textbf{333}, 996 (2011); J. Simon, W.S. Bakr, R. Ma, M.E. Tai, P.M. Preiss, and M. Greiner, 
Nature (London) \textbf{472}, 307 (2011), and references therein.

                          

\bibitem{optical-lattice}  O. Mandel, M. Greiner, A. Widera, T. Rom, T.W. H\"{a}nsch, and I. Bloch, Nature \textbf{425}, 937 (2003); I. Bloch, J. Phys. B: At. Mol. Opt. Phys. {\bf 38}, S629 (2005); P. Treutlein, T. Steinmetz, Y. Colombe, B. Lev, P. Hommelhoff, J. Reichel, M. Greiner, O. Mandel, A. Widera, T. Rom, I. Bloch, and T.W. H{\"a}nsch, Fortschr. Phys. \textbf{54}, 702 (2006); M. Cramer, A. Bernard, N. Fabbri, L. Fallani, C. Fort, S. Rosi, F. Caruso, M. Inguscio, and M.B. Plenio, Nat. Comm. \textbf{4}, 2161 (2013), and references therein.    

                             


\bibitem{solid_xy} M. Schechter and P.C.E. Stamp, Phys. Rev. B \textbf{78}, 054438 (2008), and references therein.

\bibitem{NMR} L.M.K. Vandersypen and I.L. Chuang. Rev. Mod. Phys. {\bf 76}, 1037 (2005); J. Zhang, M.H. Yung, R. Laflamme, A.A. Guzik, J. Baugh, Nat. Comm. \textbf{3}, 880 (2012); K. Rama Koteswara Rao, H. Katiyar, T.S. Mahesh, A. Sen(De), U. Sen, and A. Kumar, Phys. Rev. A \textbf{88}, 022312 (2013), and references therein.          
      
\bibitem{ksgrmp} A. Polkovnikov, K. Sengupta, A. Silva, and M. Vengalattore, Rev. Mod. Phys. \textbf{83}, 863 (2011).
                              
\bibitem{eisert_nat_phy} J. Eisert, M. Friesdorf, and  C. Gogolin, Nat. Phys.
 \textbf{11}, 124 (2015).
 



                     
\bibitem{amicoetal2004} L. Amico, A. Osterloh, F. Plastina, R. Fazio, G.M. Palma, Phys. Rev. A. \textbf{69}, 022304 (2004); L. Amico and A. Osterloh, 2004, J. Phys. A 37, 291 (2004); S.D. Hamieh, and M.I. Katsnelson,  Phys. Rev. A \textbf{72}, 032316 (2005).

\bibitem{bose2003} S. Bose, Phys. Rev. Lett. \textbf{91}, 207901 (2003); V. Subrahmanyam, Phys. Rev. A \textbf{69}, 034304 (2004); M. Christandl, N. Datta, A. Ekert, and A. Landahl, Phys. Rev. Lett. \textbf{92} 187902 (2004); D. Burgarth, S. Bose, and V. Giovannetti, Int. J. Quanum Inform. \textbf{04}, 405 (2006); Z.M. Wang, M.S. Byrd, B. Shao, and J. Zou, Phys. Lett. A \textbf{373}, 636 (2009); P.J.P. Ross and A. Kay, Phys. Rev. Lett. \textbf{106}, 020503 (2011); N.Y. Yao, L. Jiang, A.V. Gorshkov, Z.X. Gong, A. Zhai, L.M. Duan, and M.D. Lukin, Phys. Rev. Lett. \textbf{106}, 040505 (2011); H.Y.Appleby and T.J. Osborne, Phys. Rev. A \textbf{85}, 012310 (2012); T.J.G. Apollaro, S. Lorenzo, and F. Plastina, Int. J. Mod. Phys. B \textbf{27}, 1345035 (2013).


\bibitem{quenched_dynamics}  F. Igl\'{o}i and H. Rieger, Phys. Rev. Lett. \textbf{85}, 3233 (2000); K. Sengupta, S. Powell, and S. Sachdev, Phys. Rev. A \textbf{69}, 053616 (2004); S.R. Manmana, S. Wessel, R.M. Noack, and A. Muramatsu, Phys. Rev. Lett. \textbf{98}, 210405 (2007); K.R.A. Hazzard, M. van den Worm, M.F. Feig, S.R. Manmana, E.G. Dalla Torre, T. Pfau, M. Kastner, and A. Ma. Rey, Phys. Rev. A \textbf{90}, 063622  (2014); L. Bonnes, F.H.L. Essler, and A.M. L\"{a}uchli, Phys. Rev. Lett. \textbf{113}, 187203 (2014).

\bibitem{simulation_dyn} T. Kinoshita, T. Wenger, and D. Weiss, Nature \textbf{440}, 900 (2006); H. Wichterich and S. Bose, Phys. Rev. A \textbf{79}, 060302(R) (2009); Z. Chang and N. Wu, Phys. Rev. A  \textbf{81}, 022312 (2010); L. Cincio, J. Dziarmaga, M.M. Rams, and W.H. Zurek, Phys. Rev. A \textbf{75}, 052321 (2007); G. Vidal, Phys. Rev. Lett. \textbf{93}, 040502 (2004); F. Verstraete, D. Porras, and J.I. Cirac, Phys. Rev. Lett. \textbf{93}, 227205 (2004); 
S.R. Clark and D. Jaksch, Phys. Rev. A \textbf{70}, 043612 (2004); 
L. Amico, A. Osterloh, F. Plastina, R. Fazio, and G. M. Palma, Phys. Rev. A \textbf{69}, 022304 (2004); W. D\"{u}r, L. Hartmann, M. Hein, M. Lewenstein, and H.J. Briegel., Phys. Rev. Lett. \textbf{94}, 097203 (2005).  





 
   



\bibitem{ergo_group} R. Prabhu, A. Sen(De), and U. Sen, Phys. Rev. A {\bf 86} 012336 (2012); R. Prabhu, A. Sen(De), and U. Sen, Europhys. Letts. {\bf 102}, 30001 (2013); T. Chanda, T. Das, D. Sadhukhan, A. Kumar Pal, A. Sen(De), and U. Sen, Phys. Rev. A \textbf{94}, 042310 (2016); S. Roy, T. Chanda, T. Das, D. Sadhukhan, A. Sen De, U. Sen,  	arXiv:1710.11037 [quant-ph].

\bibitem{kastner2014} K.R.A. Hazzard, M. van den Worm, M.F.  Feig, S.R. Manmana, E.D. Torre, T. Pfau, M. Kastner, and A. Maria Rey, Phys. Rev. A \textbf{90}, 063622 (2014).

\bibitem{myservival} U. Mishra, D. Rakshit, and R. Prabhu, Phys. Rev. A \textbf{93}, 042322 (2016).

\bibitem{collapse_revival} H. S. Dhar, R. Ghosh, A. Sen(De), and U. Sen, Europhys. Lett. 98, 30013 (2012).

               

\bibitem{kibble} L. Cincio, J. Dziarmaga, M.M. Rams, Wojciech H. Zurek, Phys. Rev. A {\bf 75}, 052321.2007.

\bibitem{MBL} C. Gogolin, M.P. Mueller, and J. Eisert, Phys. Rev. Lett. {\bf 106},
040401 (2011); R. Nandkishore and D.A. Huse,  Annu. Rev. Condens. Matter Phys. {\bf 6}, 15 (2015).

\bibitem{Jafari} H.T. Quan, Z. Song, X.F. Liu, P. Zanardi, and C. P. Sun, Phys. Rev. Lett. \textbf{96}, 140604 (2006);
R. Jafari and H. Johannesson, Phys. Rev. Lett. \textbf{118}, 015701 (2017); R. Jafari and H. Johannesson, Phys. Rev. B \textbf{96}, 224302 (2017).




\bibitem{Rigol} M. Rigol, V. Dunjko, V. Yurovsky, and M. Olshanii, Phys. Rev. Lett. \textbf{98}, 050405 (2007).


\bibitem{Girardeau} M.D. Girardeau, Phys. Lett. \textbf{30 A}, 442 (1969); D.G. Joshi and M. Campisi, Eur. Phys. J. B \textbf{86}, 157 (2013).



\bibitem{nonintegrableergo} U. Mishra, R. Prabhu, A. Sen(De), and U. Sen, Phys. Rev. A {\bf 87}, 052318 (2013).







\bibitem{ergo_group1} A. Sen (De), U. Sen, and M. Lewenstein, Phys. Rev. A  {\bf 70} , 060304(R) (2004).

\bibitem{sabre}  Z. Huang and S. Kais, Phys. Rev. A \textbf{73}, 022339 (2006); G. Sadiek, B. Alkurtass, O. Aldossary, Phys. Rev. A \textbf{82}, 052337 (2010);
B. Alkurtass, G. Sadiek, and S. Kais, Phys. Rev. A \textbf{84}, 022314 (2011).

\bibitem{book} H.J. St{\"o}ckmann,\textit{ Quantum Chaos: An Introduction}; M. G. Grifoni and   P. H{\"a}nggi,  Phys. Rep. \textbf{304}, 229 (1998).






\bibitem{dynamic-localization} T. Prosen,  Phys.  A:  Math.  Gen., 31, L645 (1998); Phys.  Rev.  Lett. \textbf{80},  1808 (1998).

\bibitem{Rudner}  T. Kitagawa, E. Berg, M. Rudner, and E. Demler,
Phys. Rev. B \textbf{82}, 235114 (2010).

\bibitem{hopping} A. Eckardt, C. Weiss, M. Holthaus, Phys. Rev. Lett. \textbf{95}, 260404 (2005); H. Miyake, G.A. Siviloglou, C.J. Kennedy, W. C. Burton, and W. Ketterle, Phys. Rev. Lett. \textbf{111}, 185302 (2013); F. Meinert, M.J. Mark, K. Lauber, A.J. Daley, and H.C. N{\"a}gerl, Phys. Rev. Lett. \textbf{116}, 205301 (2016).

\bibitem{arti-gauge} M. Aidelsburger, M. Atala, M. Lohse, J.T. Barreiro, B. Paredes, and I. Bloch, Phys. Rev. Lett. \textbf{111}, 185301 (2013); N. Goldman and J. Dalibard
Phys. Rev. X \textbf{4}, 031027  (2014); O. Kyriienko and Anders S. SN{\o}rensen, arXiv:1703.04827 [quant-ph].



\bibitem{krishnendu2016} A. Sen, S. Nandy, and K. Sengupta, Phys. Rev. B \textbf{94}, 214301 (2016).

\bibitem{fazio2016} A. Russomanno, G.E. Santoro, R. Fazio, J. Stat. Mech. \textbf{2016}, 073101 (2016).

\bibitem{Apollaro2016} T.J.G. Apollaro, G.M. Palma, and J. Marino, Phys. Rev. B 94, 134304 (2016). 




\bibitem{heating-pd} F. Plastina, A. Alecce, T.J.G. Apollaro, G. Falcone, G. Francica, F. Galve, N. Lo Gullo, R. Zambrini, Phys. Rev. Lett. \textbf{113}, 260601 (2014).

\bibitem{pd-two-site-ent} S. Lorenzo, J. Marino, F. Plastina, G.M. Palma, T.J.G. Apollaro, Sci. Rep. \textbf{7}, 5672 (2017).

      \bibitem{Lazarides} A. Lazarides, A. Das, and R. Moessner, Phys. Rev. E \textbf{90}, 012110 (2014) and references therein. 



\bibitem{lab_xy_ion} D. Leibfried, R. Blatt, C. Monroe, and D. Wineland, Rev. Mod. Phys. {\bf 75}, 281 (2003); X.L. Deng, D. Porras, and J.I. Cirac, Phys. Rev. A \textbf{72}, 063407 (2005); H. H{\"a}ffner, C. Roos, and R. Blatt, Phys. Rep. {\bf 469}, 155 (2008); K. Kim, M.S. Chang, S. Korenblit, R. Islam, E.E. Edwards, J. K. Freericks, G.D. Lin, L.M. Duan, and C. Monroe, Nature (London) \textbf{465}, 590 (2010); R. Islam, E.E. Edwards, K. Kim, S. Korenblit, C. Noh, H. Carmichael, G.D. Lin, L.M. Duan, C.C. Joseph Wang, J.K. Freericks, and C. Monroe,  Nature Commun. {\bf2}, 377 (2011); J. Struck, M. Weinberg, C. \"{O}lschl\"{a}ger, P. Windpassinger, J. Simonet, K. Sengstock, R. H\"{o}ppner, P. Hauke, A. Eckardt, M. Lewenstein, and  L. Mathey, Nat. Phy. \textbf{9}, 738 (2013), and references therein.                                               


\bibitem{xxz-exp} J.J. Garc\'{i}a-Ripoll and J.I. Cirac, New. J. Phys. \textbf{5}, 76 (2003); L.M. Duan, E. Demler, and M.D. Lukin, Phys. Rev. Lett. \textbf{91}, 090402 (2003), and references therein. 

\bibitem{Eckardt} A. Eckardt, Rev. Mod. Phys. {\bf 89}, 011004 (2017).





\bibitem{Lieb} E. Lieb, T. Schultz, and D. Mattis,
Ann. Phys. (NY)\textbf{16}, 407 (1961);  E. Barouch, B.M. McCoy, and M. Dresden, Phys. Rev. A \textbf{2}, 1075 (1970); E. Barouch and B.M. McCoy, \textit{ibid} \textbf{3}, 786 (1971); P. Mazur, Physica \textbf{43}, 533 (1969).

\bibitem{happola} J. H\"{a}pp\"{o}l\"{a}, G. B. Hal\'{a}sz, and A. Hamma, Phys. Rev. A \textbf{85}, 032114  (2012).





\bibitem{LEV_email} L. Vidmar and M. Rigol, J. Stat. Mech. {\bf 06}, 064007 (2016).


\bibitem{refrepcit1} A. Osterloh, G. Palacios, and S. Montangero
Phys. Rev. Lett. \textbf{97}, 257201 (2006)
\bibitem{refrepcit2} B. Tomasello, D. Rossini, A. Hamma, and L. Amico,
	Europhys. Lett. 96, 27002 (2011).

\bibitem{ujjwal-review} A. Bera, T. Das, D. Sadhukhan, S.S. Roy, A. Sen De, U. Sen, 	arXiv:1703.10542 [quant-ph].










 
\end{thebibliography}
\end{document}